\documentclass[traditabstract]{aa} % for the abstract without structuration 
\usepackage{graphicx}
\usepackage{txfonts}
\begin{document}
   \title{The SOPHIE search for northern extrasolar planets \thanks{Based on observations made with 
the ELODIE and the SOPHIE spectrographs on the 1.93-m telescope at Observatoire de Haute-Provence (OHP, CNRS/OAMP), France (program 07A.PNP.CONS) and on spectral data retrieved from the ELODIE archive at OHP. }\fnmsep\thanks{Tables A.1 to A.10  are only available in electronic form at the CDS via anonymous ftp to cdsarc.u-strasbg.fr (130.79.128.5) or via http://cdsweb.u-strasbg.fr/cgi-bin/qcat?J/A+A/}}

   \subtitle{V. Follow-up of ELODIE candidates: Jupiter-analogs around Sun-like stars}

   \author{I. Boisse
          \inst{1,2}
\and F. Pepe\inst{3}
\and C. Perrier\inst{4} 
\and D. Queloz\inst{3}
\and X. Bonfils\inst{4}
\and F. Bouchy\inst{2,5}
\and N.C. Santos\inst{1,6}
\and L. Arnold\inst{5}
\and J.-L. Beuzit\inst{4}
\and R.F. D\'iaz\inst{7}
\and X. Delfosse\inst{4}
\and A. Eggenberger\inst{4}  
\and D. Ehrenreich\inst{4}
\and T. Forveille\inst{4} 
\and G. H\'ebrard\inst{2,5}
\and A.-M. Lagrange\inst{4}
\and C. Lovis\inst{3}
\and M. Mayor\inst{3}
\and C. Moutou\inst{7}
\and D. Naef\inst{3} 
\and A. Santerne\inst{7}
\and D. S\'egransan\inst{3}
\and J.-P. Sivan\inst{7}
\and S. Udry\inst{3}   
          }

   \institute{
Centro de Astrof\'isica, Universidade do Porto, Rua das Estrelas, 4150-762 Porto, Portugal\\
              \email{Isabelle.Boisse@astro.up.pt}
 \and
Institut d'Astrophysique de Paris, UMR7095 CNRS, Universit\'e Pierre \& Marie Curie, 
98bis Bd Arago, 75014 Paris, France
\and
Observatoire de Gen\`eve, Universit\'e de Gen\`eve, 51 Ch. des Maillettes, 1290 Sauverny, 
Switzerland
\and 
UJF-Grenoble 1 / CNRS-INSU, Institut de PlanŽtologie et d'Astrophysique de Grenoble (IPAG) UMR 5274, Grenoble, F-38041, France
\and
Observatoire de Haute Provence, CNRS/OAMP, 04870 St Michel l'Observatoire, France
\and
Departamento de F\'{\i}sica e Astronomia, Faculdade de Ci\^encias,
Universidade do Porto, Rua do Campo Alegre, 4169-007 Porto, Portugal
\and 
Laboratoire d'Astrophysique de Marseille, Universit\'e de Provence \& CNRS, 38 rue Fr\'ed\'eric Joliot-Curie, 
13388 Marseille cedex 13, France
        }

   \date{Received XX; accepted XX}

% \abstract{}{}{}{}{} 
% 5 {} token are mandatory
 
  \abstract
   {We present radial-velocity measurements obtained in one of a number of programs underway to search for extrasolar planets with the spectrograph SOPHIE at the 1.93-m telescope of the Haute-Provence Observatory. Targets were selected from catalogs observed with ELODIE, which had been mounted previously at the telescope, in order to detect long-period planets with an extended database close to 15 years.
   
Two new Jupiter-analog candidates are reported to orbit the bright stars HD150706 and HD222155 in 16.1~yr and 10.9~yr at 6.7$^{+4.0}_{-1.4}$~AU and 5.1$^{+0.6}_{-0.7}$~AU and to have minimum masses of 2.71$^{+1.14}_{-0.66}$~M$_{\mathrm Jup}$ and 1.90$^{+0.67}_{-0.53}$~M$_{\mathrm Jup}$, respectively. Using the measurements from ELODIE and SOPHIE, we refine the parameters of the long-period planets HD154345b and HD89307b, and publish the first reliable orbit for HD24040b. This last companion has a minimum mass of 4.01\,$\pm$\,0.49~M$_{\mathrm Jup}$ orbiting its star in 10.0~yr at 4.92\,$\pm$\,0.38~AU. Moreover, the data provide evidence of a third bound object in the HD24040 system. 
   
   With a surrounding dust debris disk, HD150706 is an active G0 dwarf for which we partially corrected the effect of the stellar spot on the SOPHIE radial-velocities. In contrast, HD222155 is an inactive G2V star. In the SOPHIE measurements, an instrumental effect could be characterized and partly corrected.  On the basis of the previous findings of Lovis and collaborators and since no significant correlation between the radial-velocity variations and the activity index are found in the SOPHIE data, these variations are not expected to be only due to stellar magnetic cycles.
   Finally, we discuss the main properties of this new population of long-period Jupiter-mass planets, which for the moment, consists of fewer than 20 candidates.
      These stars are preferential targets either for direct-imaging or astrometry follow-up surveys to constrain the system parameters and for higher-precision radial-velocity searches for lower mass planets, aiming to find a solar system twin.
   In the Appendix, we determine the relation that defines the radial-velocity offset between the ELODIE and SOPHIE spectrographs.

 }


   \maketitle
%
%________________________________________________________________

\section{Introduction}

One motivation of our search for planetary systems is to put  the solar system into perspective and understand its formation. Until now, most discovered systems have not resembled the planets of our system. If we were to observe the Sun in radial-velocity (hereafter RV), the main source of perturbation should be that of Jupiter with a period of 11.86 yr, an orbital distance of 5.2 UA, and a RV semi-amplitude of about 12 ms$^{-1}$. The detection of long-period Jupiter-like planets are therefore expected to be the first step in the quest to discover an analog of the solar system. This step is already achievable, in contrast to the detection of earth-like planets in the habitable zone of their host stars which will generally require the next generation of instruments.

The long-term accuracy of several spectrographs and the timescale of some RV surveys has started to permit the discoveries of long-period planets. Few planets are known to date within the orbital distance range of Jupiter with fewer than seventeen planets having been found to be orbiting at distances greater than 4~AU (cf. Table~\ref{liste}). Some of these have been announced with incomplete orbits.  
These planets overlap with the few microlensing detections at these orbital distances and in the Jupiter-mass regime, as OGLE235-MOA53b is a 2.6~M$_{Jup}$ planet at 5.1AU (Bond et al. 2008). We note that the planet-host stars are expected to be low-mass stars and their giant planets are colder than Jupiter. %p.9 du cahier: quelles sont les autres plantes ?

%--------------------------------------
\begin{table*}[h]
  \centering 
  \caption{Known exoplanets discovered by RV with orbital distances greater than 4~AU.}
  \label{liste}
\begin{tabular}{lccccccc}
\hline
\hline
                          & Semi-major axis[AU] & Period [day] & Mass [M$_{\mathrm Jup}$] & Ecc  &  Orbit$^{\star}$ & Multiplicity$^{\dag}$  & ref. \\
\hline
HIP70849\,b        &  4.5-36                     & 5-90[yr]                       & 3-15                     & 0.47-0.96           &  incomp.                & 1       & S\'egransan et al. (2011)\\
HD150706\,b        &  6.7$^{+4.0}_{-1.4}$~$^{\star\star}$  & 5894$^{+5584}_{-1498}$  & 2.71$^{+1.14}_{-0.66}$~$^{\star\star}$             &  0.38$^{+0.28}_{-0.32}$  & incomp.                   & 1        & this paper \\
HD134987\,c         &  5.8       & 5000$\pm$400  & 0.82$\pm$0.03        & 0.12$\pm$0.02      & incomp.                 &    2                  & Jones et al. (2010) \\
55Cnc\,d               & 5.76              &  5218$\pm$230   & 3.84$\pm$0.08     &    0.03$\pm$0.03  & comp.                 &       5                    &  Fischer et al. (2008) \\
HD190984\,b         &   5.5    & 4885$\pm$1600    &  3.1                      & 0.57$\pm$0.10      & incomp.                  & 1                    &   Santos et al. (2010b) \\
HD99492\,c         &   5.4             &    4970$\pm$744  & 0.36$\pm$0.06      &      0.1$\pm$0.2    & comp.                  &  2           & Meschiari et al. (2010) \\
HD217107\,c         &  5.27     &  4270$\pm$220   &2.60$\pm$0.15         & 0.517$\pm$0.033  & incomp.                &     2                  & Wright et al. (2009) \\
$\mu$ Ara\,e    &    5.24               &   4206$\pm$759      & 1.8                   &   0.10$\pm$0.06        & comp.                  &        4           & Pepe et al. (2007) \\
HD13931\,b         &   5.15              &      4218$\pm$388 & 1.88$\pm$0.15      &     0.02$\pm$0.05   & comp.                  &      1            &  Howard et al. (2010) \\
HD222155\,b        & 5.1$^{+0.6}_{-0.7}$~$^{\ast\ast}$    & 3999$^{+469}_{-541}$  & 1.90$^{+0.67}_{-0.53}$~$^{\ast\ast}$               & 0.16$^{+0.27}_{-0.22}$    & comp.                   & 1                      & this paper \\
HD7449\,c            & 4.96$\pm$0.30  & 4046$\pm$370   &  2.0$\pm$0.5                & 0.53$\pm$0.08       & comp.     & 2         & Dumusque et al. (2011) \\
HD24040\,b          & 4.92$\pm$0.38$^{\bigtriangleup}$    & 3668$^{+169}_{-171}$ & 4.01$\pm$0.49$^{\bigtriangleup}$                   &    0.04$^{+0.07}_{-0.06}$ & comp.                 &      1(+1?$^{\ast}$)    & this paper \\
HD220773\,b        &  4.94$\pm$0.2      & 3725$\pm$463    & 1.45$\pm$0.3    & 0.51$\pm$0.1        & incomp. & 1 & Robertson et al. (2012)\\
HD187123\,c         & 4.89$\pm$0.53    &  3810$\pm$420    & 1.99$\pm$0.25  & 0.252$\pm$0.033   & comp. & 2 & Wright et al. (2009) \\
HD72659\,b          & 4.74$\pm$0.08    &  3658$\pm$32     & 3.15$\pm$0.14  & 0.22$\pm$0.03    &  comp. &    1    & Moutou et al. (2011)  \\
HD183263\,c         & 4.35$\pm$0.28     &  3070$\pm$110   &  3.57$\pm$ 0.55 & 0.239$\pm$0.064    & incomp.  & 2 & Wright et al. (2009) \\
HD106270\,b        & 4.3$\pm$0.4      &  2890$\pm$390    &  11$\pm$1   & 0.40$\pm$0.05    & incomp.  & 1 & Johnson et al. (2011) \\
HD154345\,b        & 4.3$\pm$0.4$^{\bigtriangleup\bigtriangleup}$   & 3538$\pm$300 &   1.0$\pm$0.3$^{\bigtriangleup\bigtriangleup}$               &   0.26$\pm$0.15 & comp.                   &     1                      & this paper \\
HD125612\,d          & 4.2       & 3008$\pm$202   &  7.2   & 0.28$\pm$0.12    & incomp.  &  3$^{\ddag}$ & Lo Curto et al. (2010) \\  \hline

\end{tabular}
\begin{list}{}{}
\item[$^{\star}$] \textbf{comp.}: if the RV measurements cover the orbit; \textbf{incomp.}: if the RV measurements did not cover the orbit.
\item[$^{\dag}$] Give the number of planets \textbf{pl.} in the system. All planets, except HD125612b, orbit single stars. 
\item[$^{\ast}$] A linear trend is fit to the RV of this system. 
\item[$^{\ddag}$] HD125612A has a stellar companion
\item[$^{\star\star}$] Assuming M$_{\star}$\,=\,1.17\,$\pm$\,0.12\,M$_{\odot}$ 
\item[$^{\ast\ast}$] Assuming M$_{\star}$\,=\,1.13\,$\pm$\,0.11\,M$_{\odot}$
\item[$^{\bigtriangleup}$] Assuming M$_{\star}$\,=\,1.18\,$\pm$\,0.10\,M$_{\odot}$
\item[$^{\bigtriangleup\bigtriangleup}$] Assuming M$_{\star}$\,=\,0.88\,$\pm$\,0.09\,M$_{\odot}$

\end{list}

\end{table*}

%---------------------------------------------------

The SOPHIE consortium started a large program to search for planets in October 2006 (Bouchy et al. 2009) that led to several planet discoveries (e.g. H\'ebrard et al. 2010, Boisse et al. 2010b, D\'iaz et al. 2012). Among the different subprograms, one focuses on the follow-up of the drifts and long-period signals detected in the ELODIE sample, in line with the continuity of the historical program initiated by M. Mayor and D. Queloz in 1994 with the spectrograph ELODIE over more than 12 years (Mayor \& Queloz 1995; Naef et al. 2005). These trends are identified as incomplete orbits of gravitationally bound companions and the monitoring aims to determine their periods and masses, (thus establish either their planetary, brown dwarf, or stellar nature). The SOPHIE spectrograph has replaced ELODIE at the 1.93m-telescope at Observatoire de Haute-Provence (OHP) since October 2006. About 40 targets were selected from the original ELODIE catalog, which contained about 400 targets. They are mainly G and K dwarfs, which have been observed with SOPHIE with the objective of detecting very long-period planets ($>$8 yr) and multiple systems. 

We report the detection of two Jupiter-analogs around the Sun-like stars HD150706 and HD222155 based on ELODIE and SOPHIE RV measurements.
The observations are presented in Section 2, and we characterize the planet-host stars in Sect.~3. In Sect.~4, we analyze the RV measurements and constrain the planetary parameters. In Sect.~5, we determine the first reliable orbit for HD24040b (Wright et al. 2007), and refine the planetary parameters of HD154345b (Wright et al. 2008) and HD89307b (Fischer et al. 2009). Finally in Sect.~6, we discuss how the observed RV variations should not come from long-term magnetic cycles, before putting these new planets in the context of the other discoveries and the perspectives for these systems to be followed. In the Appendix, we determine the RV shift between the ELODIE and SOPHIE data.

%__________________________________________________________________

\section{Radial velocity measurements}
\label{obs}
Measurements were obtained with the cross-dispersed echelle ELODIE spectrograph mounted on the 1.93-m telescope at the Observatoire de Haute-Provence observatory (OHP, France) between late 1993 and mid 2006 (Baranne et al. 1996). 
The stars were subsequently monitored by the SOPHIE spectrograph that replaced ELODIE which provided improvements in terms of stability, limiting magnitude, and resolution. 
For the two instruments, the stellar spectrum were recorded simultaneously with a thorium-argon calibration (thosimult mode, Bouchy et al. 2009), allowing an estimation of the intrinsic drift of the spectrograph at the same time as the observation. The optical fibers include a double scrambler in the path of light to improve the RV stability. A mean time exposure of 900~s (which varied between 600~s and 1200s depending on the weather conditions) helped to minimize the photon noise and average the acoustic oscillation modes (p-modes).  

ELODIE has a resolving power R=$\lambda/\Delta\lambda\,\approx\,42\,000$  (at 550~nm, see e.g. Perrier et al. 2003 for more details). The spectra are correlated with a G2-spectral type numerical mask. The resulting cross-correlation functions (CCF) are fitted by Gaussians to derive the RV values (Baranne et al. 1996, Pepe et al. 2002).

The SOPHIE observations were performed in high-resolution mode reaching a resolution power of $\Delta\lambda/\lambda\,\approx\,75\,000$ (at 550~nm). The SOPHIE automatic data reduction software was used to derive the RV from the spectra, after a cross-correlation with a G2-spectral type numerical mask (Baranne et al. 1996; Pepe et al. 2002) and a fit of a Gaussian to the resulting CCF. The typical photon-noise uncertainty is around 1.5~ms$^{-1}$, which was calculated as described in Boisse et al. 2010b. However, the main error source in these measurements originates from the instrument, namely the \textit{seeing effect} (Boisse et al. 2010a,b). 
This instrumental effect is due to the insufficient scrambling of one multimode fiber that leads to the non-uniform illumination of the entrance of the spectrograph. We note that this noise was removed by a fiber link modification, which includes a piece of octagonal-section fiber in June 2011 (Perruchot et al. 2011).
 An external systematic error of 4~ms$^{-1}$ for instrumental errors (guiding, centering, and seeing) was then quadratically added to the SOPHIE mean measurement uncertainty. Seeing error is not expected in the ELODIE measurements as the instrumental configuration was different, but the RV uncertainty take into account the guiding and centering errors. In the following, the signal-to-noise ratio (SNR) are given per pixel at 550~nm. We note the sampling per resolution element (full width half maximum, FWHM) is 2.2 pixels for ELODIE and 2.7 pixels in the high-resolution mode for SOPHIE.

The RV data are available at the CDS as tables, which contain in theirs cols. 1-3, the time of the observation (barycentric Julian date), the RV, and its error, respectively

	\subsection{HD150706}

Over nine years, between July 1997 and June 2006, 50 RV measurements were done with the ELODIE spectrograph. We did not take into account two measurements with SNR$<$10. SOPHIE obtained 59 observations of HD150706 between May 2007 and April 2011. Five measurements with SNR lower than 100 were removed. One spectrum contaminated by moonlight was also discarded. The final data set contains 48 ELODIE and 53 SOPHIE measurements with a typical SNR of, respectively, 80 and 172.  The data are available at the CDS in Tables~\ref{table_rv} (ELODIE) and~\ref{table_rv2} (SOPHIE).

	\subsection{HD\,222155}

We obtained 44 spectra of HD222155 with the ELODIE spectrograph on a timescale of eight years between August 1997 and November 2005. 
HD222155 was observed 71 times by SOPHIE between July 2007 and January 2011. Three measurements with SNR lower than 100 were removed. We discarded one observation, which had been made with the background sky spectrum recorded simultaneously (\textit{objAB} mode, Bouchy et al. 2009) in order to measure the stellar parameters (cf.~Sect.~\ref{star_222}). The final data set comprises 44 ELODIE measurements with a typical SNR of 92 and 67 SOPHIE values with a mean SNR of 173. The data are available at the CDS in Tables~\ref{table_rv3} (ELODIE) and~\ref{table_rv4} (SOPHIE).   
%__________________________________________________________________

\section{Planet host stars}
% spectral type determination: cf.p. 22 du cahier

	\subsection{HD150706}
	\label{150_star}
HD150706 (HIP80902) is a G0V star with an apparent Johnson V-band magnitude of m$_\mathrm{V}$=7.0 (Hipparcos catalogue, ESA 1997). Assuming an astrometric parallax of $\pi$=35.43$\pm$0.33\,mas (van Leeuwen 2007), we derived a distance of 28.2$\pm$0.3~pc, which leads to an absolute V-band magnitude of 4.75. A stellar diameter R$_{\star}$=0.96$\pm$0.02\,R$_{\odot}$ was estimated by Masana et al. (2006) from photometric measurements.	                    
               
A spectroscopic analysis (Santos et al. 2004) was done on high-resolution spectra obtained with the UES spectrograph on the 4-m William Herschel telescope (Santos et al. 2003). They derived an effective temperature T$_\mathrm{eff}$=5961\,$\pm$\,27~K, a surface gravity log g=4.5\,$\pm$\,0.1, a metallicity [Fe/H]\,=\,-0.01\,$\pm$\,0.04, and a stellar mass M$_{\star}$\,=\,1.17\,$\pm$\,0.12\,M$_{\odot}$.

From the SOPHIE CCF (Boisse et al. 2010b), we estimated $v \sin i_{\star}$=3.7\,$\pm$\,1.0 kms$^{-1}$ and a metallicity of [Fe/H]=0.08\,$\pm$\,0.10 in agreement with the spectroscopic analysis.
% param stellaire, notes cahier p.1
We assessed the stellar activity level from the emission in the core of the \ion{Ca}{II}~H\&K bands, which was measured in each SOPHIE spectra of HD150706 with the calibration reported in Boisse et al. (2010b). This yields a value of $\log$R$^{'}_\mathrm{HK}$=-4.47\,$\pm$\,0.10. HD150706 is an active star and we may expect to find a RV jitter caused by stellar spots of about 15\,ms$^{-1}$ (Santos et al. 2000). %note cahier p.1
According to the calibrations of Noyes et al. (1984) and Mamajek \& Hillenbrand (2008), 
the $\log$\,$R'_\mathrm{HK}$ value of HD\,150706 implies a rotation period of 
P$_{rot}$\,$\approx$\,5.6 days. From the $v \sin i_{\star}$ and stellar radius values, we inferred that P$_{rot}$ $\leqslant$ 18
days (Bouchy et al. 2005). 

Holmberg et al. (2009) estimated an age of 5.1$^{+3.7}_{-4.5}$~Gyr, which agrees with the Marsakov et al. (1995) value of 4.69~Gyr. However younger ages were derived from \ion{Ca}{II} measurements, namely  1.4~Gyr (Wright et al. 2004) and 1.16~Gyr (Rocha-Pinto et al. 2004), and by comparing with stellar isochrones, namely 2.3~Gyr (Gonzalez et al. 2010). 

 Meyer et al. (2004) detected a dust debris disk surrounding HD150706 using IRAC and MIPS \textit{Spitzer} data. The authors argued for the presence of a companion in order to explain their observation of a large inner hole in the dust distribution of the disk.

The parameter values for the star are gathered in Table~\ref{param_star}.

%---------------------------
\begin{table}
  \centering 
  \caption{Stellar parameters for HD\,150706 and HD\,222155. }
  \label{param_star}
\begin{tabular}{lcc}
\hline
\hline
Parameters   & HD\,150706 & HD\,222155 \\
\hline
Sp. T.              &   G0V  &    G2V \\
m$_{V}$             &  7.0  &   7.1  \\
B - V                &  0.57  &   0.64    \\
$\pi$     [mas]  &   35.43\,$\pm$\,0.33   &    20.38\,$\pm$\,0.62   \\
T$_\mathrm{eff}$    [K]  & 5961\,$\pm$\,27$^{(1)}$      &   5765\,$\pm$\,22 $^{(2)}$     \\
log g    [cgs]  &    4.5\,$\pm$\,0.1 $^{(1)}$   &    4.10\,$\pm$\,0.13 $^{(2)}$   \\
Fe/H  [dex]  &   -0.01\,$\pm$\,0.04$^{(1)}$    &  -0.11\,$\pm$\,0.05 $^{(2)}$     \\
$v\sin i_{\star}$    [\,km\,s$^{-1}$]  & 3.7\,$\pm$\,1.0 $^{(3)}$    &    3.2\,$\pm$\,1.0$^{(3)}$    \\
M$_{\star}$    [M$_{\odot}$]  & 1.17\,$\pm$\,0.12$^{(1)}$     &   1.13\,$\pm$\,0.11     \\
R$_{\star}$    [R$_{\odot}$]  & 0.96\,$\pm$\,0.02    &     1.67$\pm$0.07    \\
$\log$\,$R'_\mathrm{HK}$   [dex]  &  -4.47\,$\pm$\,0.10$^{(3)}$   &   -5.06\,$\pm$\,0.10$^{(3)}$    \\
Age   [Gyr] &  1-5   &    8.2\,$\pm$\,0.7   \\
Distance  [pc] & 28.2\,$\pm$\,0.3    &  49.1\,$\pm$\,1.5     \\
\hline
\end{tabular}
\begin{list}{}{}
\item[$^{(1)}$] Parameter derived from the UES spectra (Santos et al. 2003).
\item[$^{(2)}$] Parameter derived from the SOPHIE spectrum (Santos et al. 2004).
\item[$^{(3)}$] Parameter derived from the SOPHIE CCF (Boisse et al. 2010b).

\end{list}
\end{table}
%-----------------------------------------------

	\subsection{HD\,222155}
	\label{star_222}	
% param stellaire, notes cahier p.2	
HD\,222155 (HIP116616) is a G2V bright star with an apparent Johnson V-band magnitude of m$_{V}$=7.1 (Hipparcos catalogue, ESA 1997) and a $B-V$=0.64. Van Leeuwen (2007) derived from the Hipparcos measurements a parallax of 20.38\,$\pm$\,0.62\,mas, leading us to infer a distance of 49.1\,$\pm$1.5\,pc with an absolute V magnitude of 3.65 mag. 

The star's effective temperature T$_{\rm eff}$\,=\,5765\,$\pm$\,22\,K, surface gravity log\,$g$\,=\,4.10\,$\pm$\,0.13, micro-turbulence velocity V$_{t}$\,=\,1.22\,$\pm$\,0.02\,\,km\,s$^{-1}$, and metallicity [Fe/H]\,=\,-0.11\,$\pm$\,0.05 dex, were determined using the spectroscopic analysis method described in Santos et al. (2004). The analysis was performed on a spectrum of high SNR measured with SOPHIE without a simultaneous calibration. When combined with isochrones (da Silva et al. 2006)\footnote{Web interface available on http://stev.oapd.inaf.it/cgi-bin/param.}, these parameters 
yield a stellar mass M$_{\star}$\,=\,1.06\,$\pm$\,0.10\,M$_{\odot}$, a stellar radius of R$_{\star}$=1.67$\pm$0.07\,R$_{\odot}$, and an age 8.2$\pm$0.7~Gyr in agreement with the 8.4~Gyr estimated by Holmberg et al. (2009). The values of the radius and the mass agree with those derived by Allende-Prieto \& Lambert (1999) of M$_{\star}$\,=\,1.20\,$\pm$\,0.11\,M$_{\odot}$ and R$_{\star}$=1.66$\pm$0.07\,R$_{\odot}$. For the stellar mass, we chose the mean value M$_{\star}$\,=\,1.13\,$\pm$\,0.11\,M$_{\odot}$.

The projected 
rotational velocity $v \sin i_{\star}$\,=\,3.2\,$\pm$\,1.0\,km\,s$^{-1}$ is estimated from the SOPHIE CCF (Boisse et al. 2010b). The alternative estimate of the stellar metallicity [Fe/H]\,=\,0.02\,$\pm$\,0.10 from the CCF is consistent with the more accurate determination based on spectral analysis. The stellar activity index is derived from the stellar spectra calculated in the \ion{Ca}{II}~H\&K lines, $\log$\,$R'_\mathrm{HK}$=-5.06\,$\pm$\,0.10. HD222155 is on its way to be a subgiant. These stars have a lower $\log$\,$R'_\mathrm{HK}$ (Wright et al. 2004, Lovis et al. 2011b) owing to their higher luminosities and/or lower surface gravities compared to main-sequence stars of the same color. These stars are expected to have smaller long-term variabilities than main-sequence stars. HD222155 is then a low-active star for which we expect intrinsic variability at a lower level than those caused by instrumental effects. For $\log$\,$R'_\mathrm{HK}$\,$\leqslant$\,-5.0, Mamajek \& Hillenbrand (2008) noted that the correlation between $\log$\,$R'_\mathrm{HK}$ and the Rossby number is poor and they were unable to derive a reliable relation to derive a P$_{rot}$.

The stellar parameters are given in Table~\ref{param_star}.

\section{Radial velocity analysis and planetary parameters}

	\subsection{HD150706b, a Jupiter analog around an active star}
	
	A Jupiter-mass planet in an eccentric 265-day orbit, HD150706b, was announced during the "Scientific Frontiers in Research on Extrasolar Planets" conference, Washington, in June 2002 based on ELODIE RV measurements. However, later observations led the conclusion that the RV variations are instead caused by a longer-period planet (S. Udry, private communication).
	
	Using Eq.~\ref{G22} from our Appendix~\ref{RVES}, we first fixed the $\Delta$(RV)$_{\mathrm E-S}$ between ELODIE and SOPHIE data and computed the weighted and the generalized Lomb Scargle periodograms. For both, the highest peak is detected close to 5000 days with a false alarm probability (fap)$<$0.001. The fap was generated using both Monte Carlo simulations to draw new measurements according to their error bars, and the random permutation of the date of the observations, as described in Lovis et al. (2011a). 
		
	Eight Keck measurements were published by Moro-Mart\'in et al. (2007) which showed that the short period solution was incorrect. We added these measurements to our RV data. An error of 5ms$^{-1}$ was quadratically added to their instrumental error bars in order to take into account the stellar activity jitter. The RV data were then fitted with a Keplerian model using a Levenberg-Marquardt algorithm, after selecting starting values with a genetic algorithm (S\'egransan et al. 2011). 	The $\Delta$(RV)$_{\mathrm E-S}$ was allowed to vary and the fitted value,$ -31.1$$\pm$13.6ms$^{-1}$ is in agreement with the calibration value of $\Delta$(RV)$_{E-S}$\,=\,-40\,$\pm$\,23ms$^{-1}$. The best-fit solution is consistent with an orbital period of $P$=3950 days and a semi-amplitude $K$=31\,ms$^{-1}$. 
	The residuals of the best-fit Keplerian model are equal to $\sigma_{(O-C)}$=19.5ms$^{-1}$, which consists of components of 18ms$^{-1}$ for the ELODIE RV, and 20ms$^{-1}$ for the SOPHIE ones. These values are large compared to the mean error bars. The 6.1ms$^{-1}$ dispersion for the residuals of the Keck data points may be smaller due to a small number of points and the free offset between datasets. 
	
	HD150706 is an active star and we may expect to measure some RV jitter as discussed in Sect.~\ref{150_star}. We note that by examining at the periodogram of the ($O-C$) values, a peak close to 10 days is scarcely detected, value in the domain of the $P_{\mathrm rot}$ that we derive in Sect.\ref{150_star}. With a $vsini$\,=\,3.7\,kms$^{-1}$, an anti-correlation between ($O-C$) and the bisector span (BIS) is expected if RV variations are due to stellar activity. The ELODIE measurements have an error bars of about 10ms$^{-1}$ for the RV and 20ms$^{-1}$ for the BIS. This precision hampers the detection of a correlation for data with a dispersion of 16ms$^{-1}$. On the other hand, an anti-correlation is observed in the SOPHIE data as shown in Fig.~\ref{BisOMCHD150706}. The correlation coefficient is equal to -0.56 with a fap$<$10$^{-5}$ and the Spearman coefficient is -0.47. The fap is calculated with random permutations of the RV data. As in Melo et al. (2007) and Boisse et al. (2009), we corrected the SOPHIE RV for this trend $RV_{\mathrm corrected1}$ [kms$^{-1}$]=$RV$ [kms$^{-1}$]+1.32$\times$ $BIS$ [kms$^{-1}$].
	
%-----------------------------------------------------------
   \begin{figure}
   \centering
   \includegraphics[width=7.5cm]{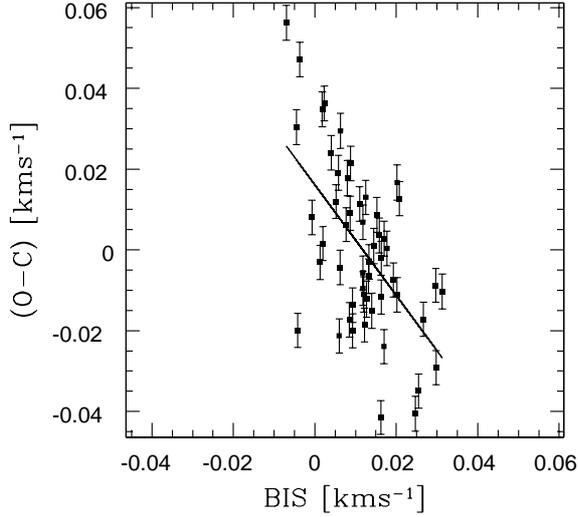}
      \caption{ SOPHIE residuals from the Keplerian fit of HD150706 as a function of the BIS. The best-linear fit is plotted as a black line. The scale is the same in the $x$ and $y$ axis.      
              }
         \label{BisOMCHD150706}
   \end{figure}
%______________________________________________________________	

	Moreover, at high SNR, SOPHIE data are polluted by an instrumental limitation, called the \textit{seeing effect} (Boisse et al. 2010a,b). This apparent RV shift is related to the illumination of the spectrograph, which varies mainly owing to the seeing. Its current characteristic signature is a linear correlation between the RV and a \textit{seeing estimator} $\ Sigma$, which accounts for the flux entering into the spectrograph per unit of time, $\Sigma\,=\,SNR^{2}/T_\mathrm{exp}$, where $T_\mathrm{exp}$ is the time exposure. The HD150706 SOPHIE $(O-C)$ are plotted in Fig.~\ref{BisOMCHD150706}. A linear trend is detected with  correlation and Spearman coefficients of -0.48 with fap$<$10$^{-4}$. 
The SOPHIE RV was corrected for this slope, $RV_{\mathrm corrected2}$ [kms$^{-1}$]=$RV_{\mathrm corrected1}$ [kms$^{-1}$]+0.00071$\times \Sigma$.	  We note that swapping the order of the corrections do not change the final result as the order of magnitude of the two effects are equivalent.		
		
%-----------------------------------------------------------
   \begin{figure}
   \centering
   \includegraphics[width=7.5cm]{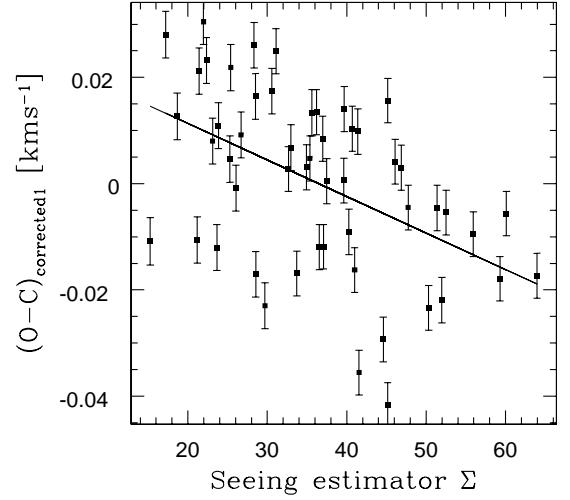}
      \caption{    HD150706 SOPHIE residuals from the Keplerian fit of the RV corrected for the active jitter as a function of the \textit{seeing estimator}, illustrating the instrumental effect on RV caused by seeing variations. The best-least squares linear fit is also plotted.           }
         \label{BisOMCHD150706}
   \end{figure}
%______________________________________________________________	

\begin{figure}
   \centering
    \includegraphics[width=8cm]{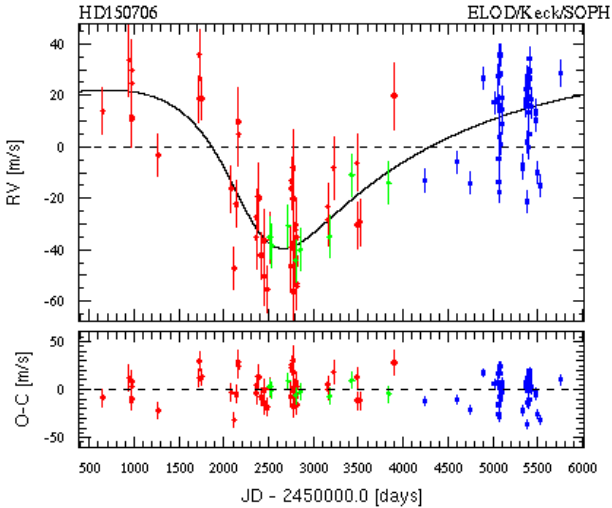}
   \caption{ ELODIE (red), Keck (green), and SOPHIE (blue) RV data points and their residuals from the best-fit Keplerian model for HD150706 as a function of barycentric Julian date. The best-fit Keplerian model is represented by the black curve with a reduced $\chi^{2}$ equal to 2.6. The period is 16.1~yr with a slight eccentricity $e$=0.38$^{+0.28}_{-0.32}$ and the planet minimum mass is 2.71~M$_{Jup}$. }
         \label{fitHD150706}
   \end{figure}

\begin{figure*}
   \centering
    \includegraphics[width=4.5cm]{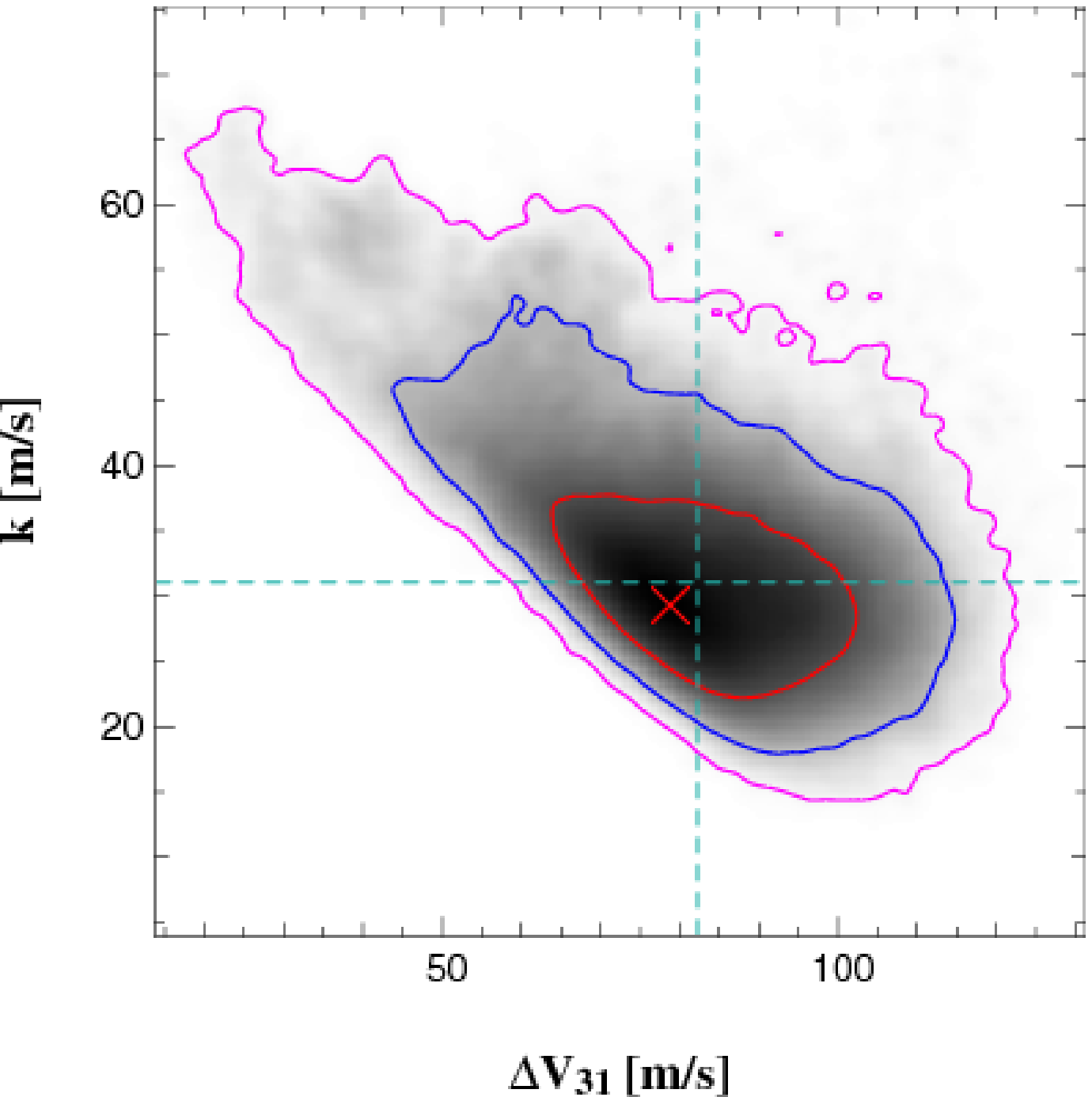}
    \includegraphics[width=4.5cm]{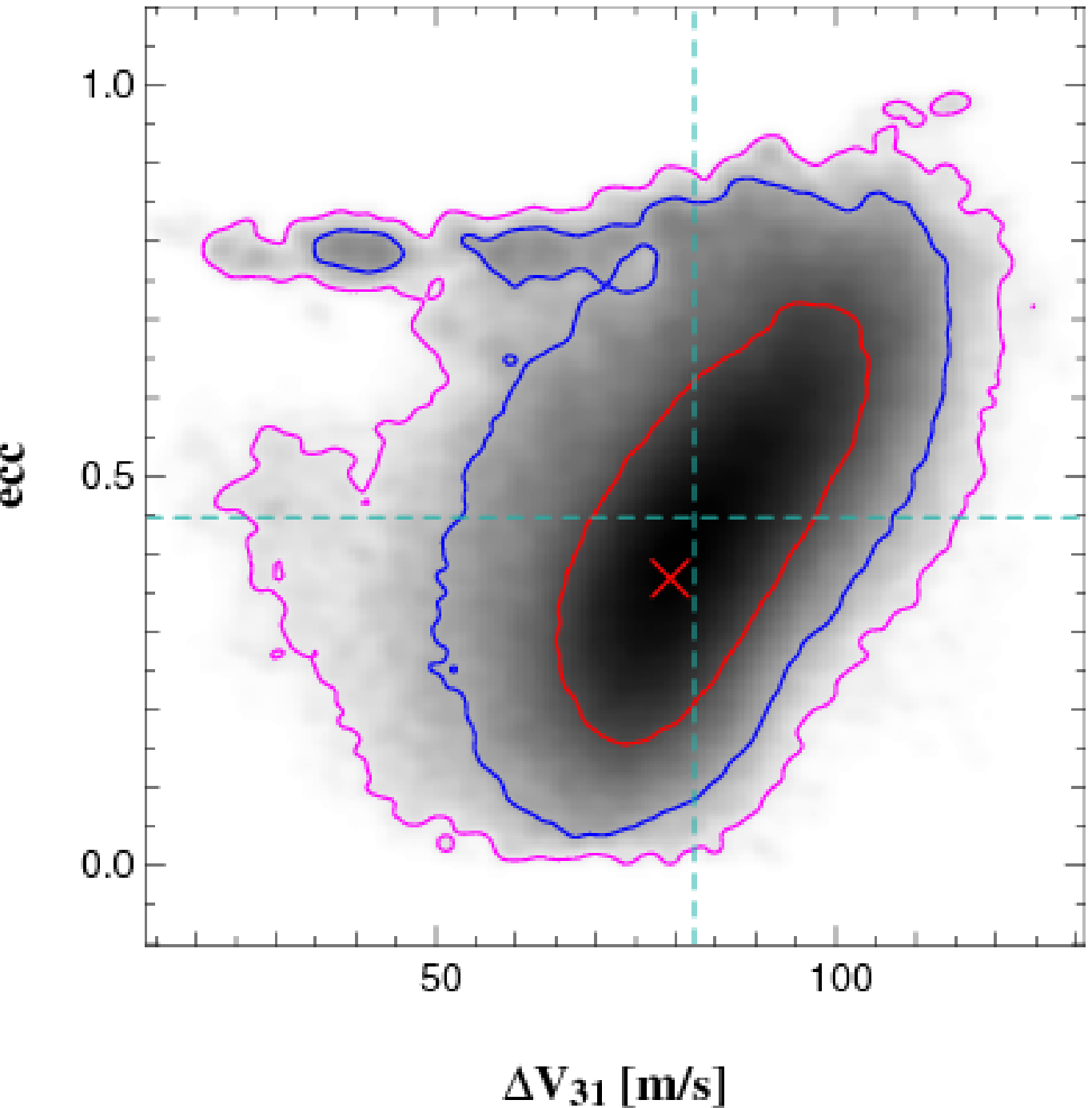}
   \includegraphics[width=6cm]{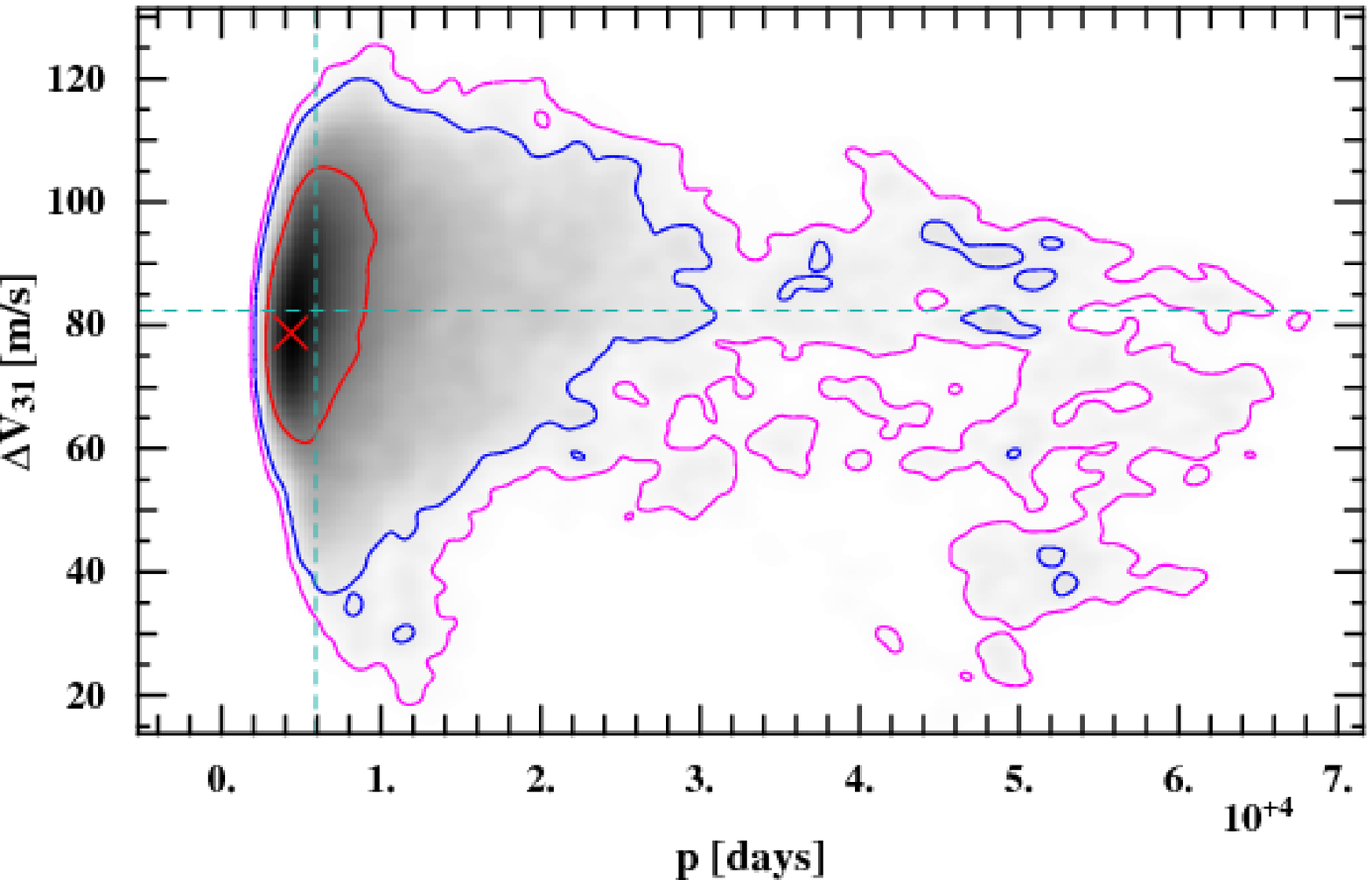}
     \includegraphics[width=4.5cm]{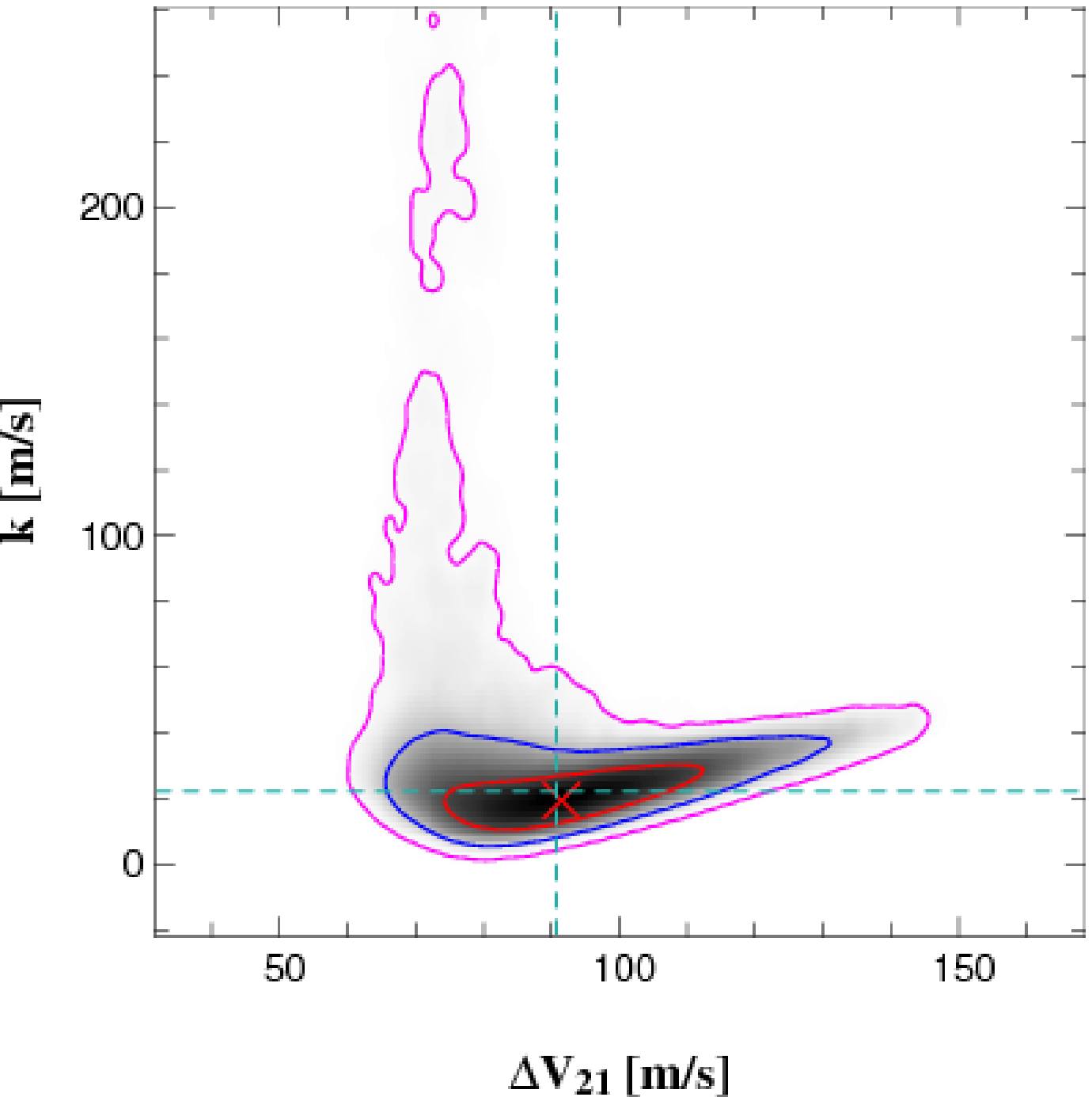}
    \includegraphics[width=4.5cm]{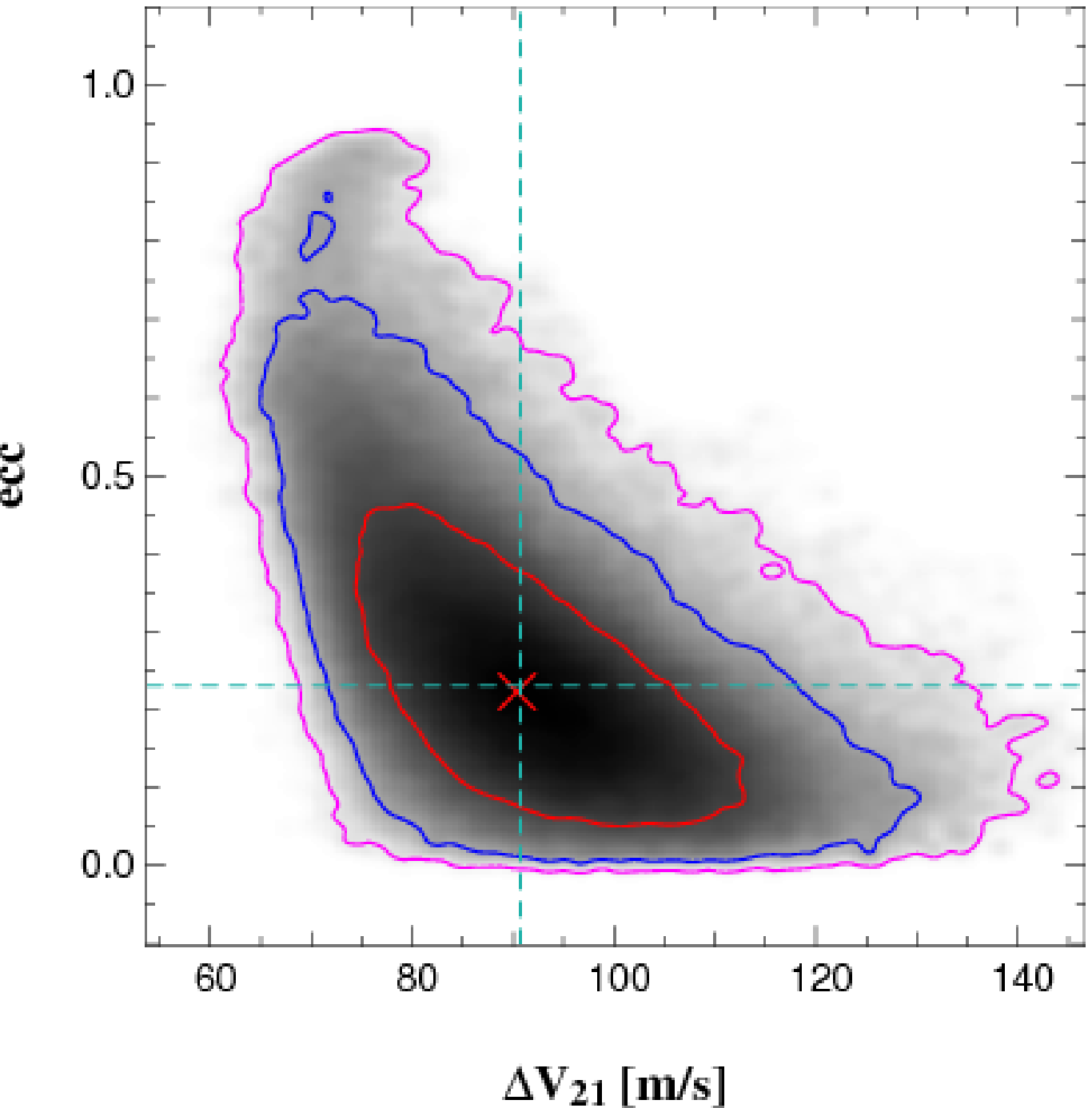}
   \includegraphics[width=6cm]{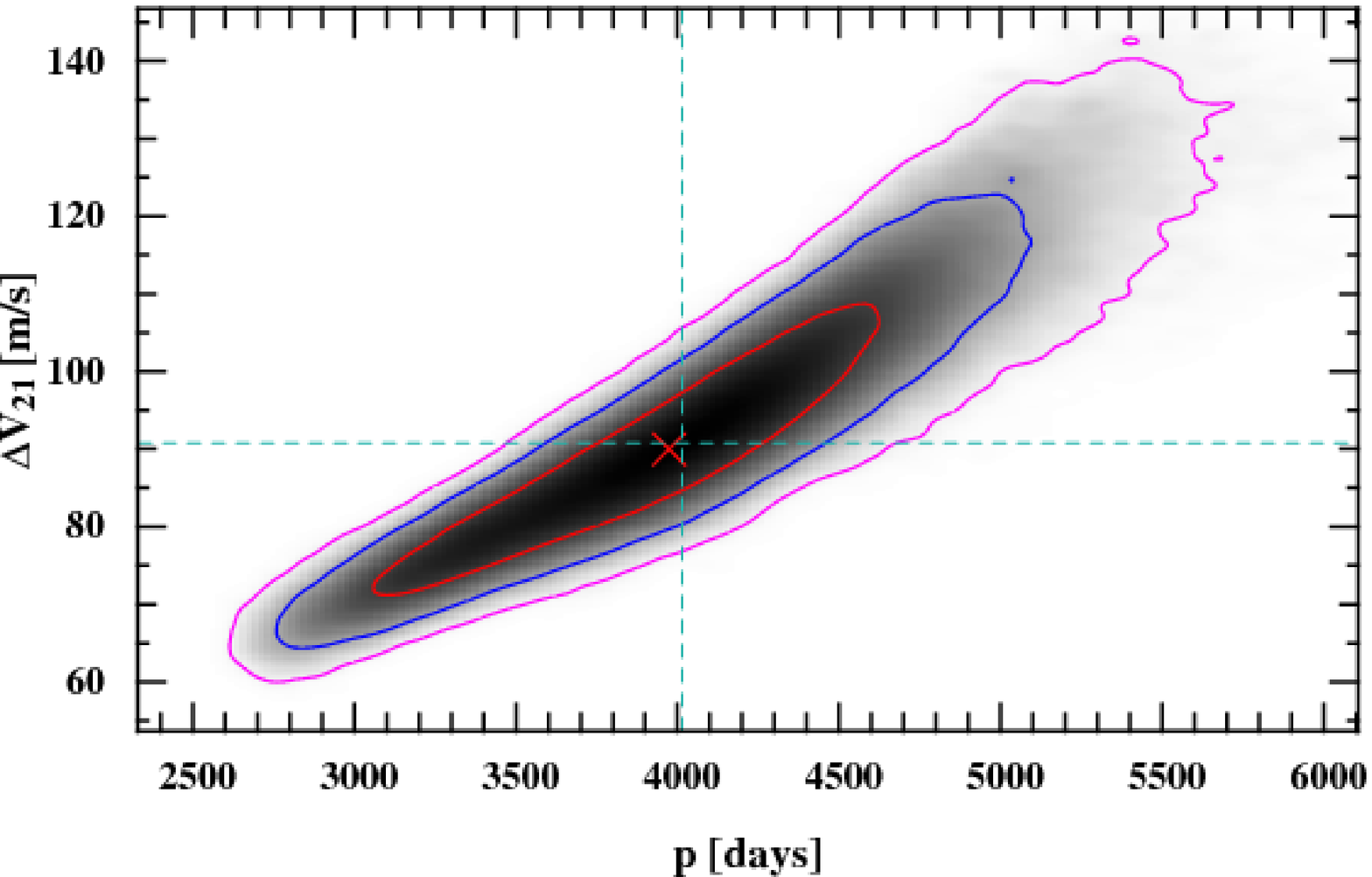}
  \includegraphics[width=4.5cm]{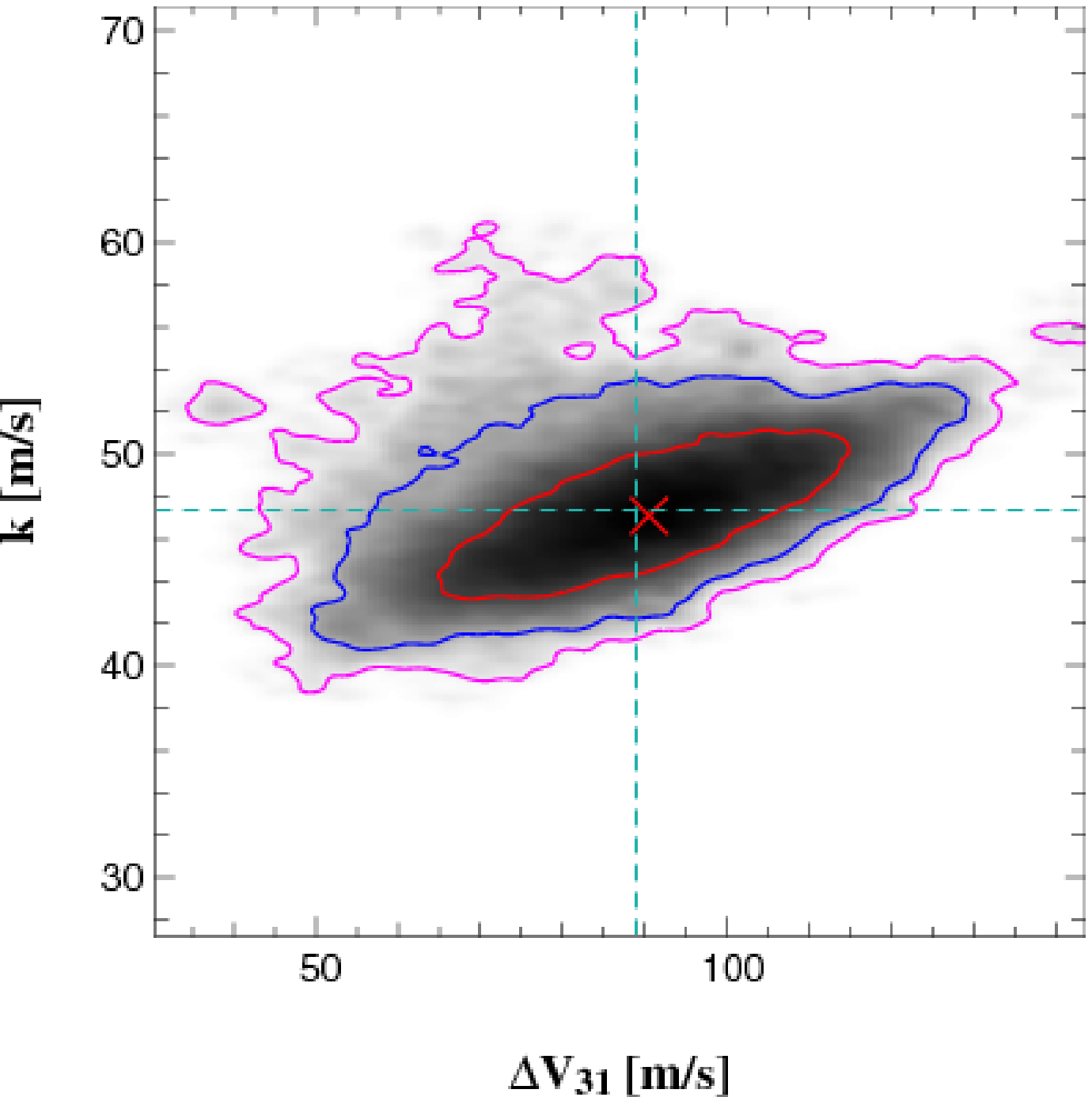}
    \includegraphics[width=4.5cm]{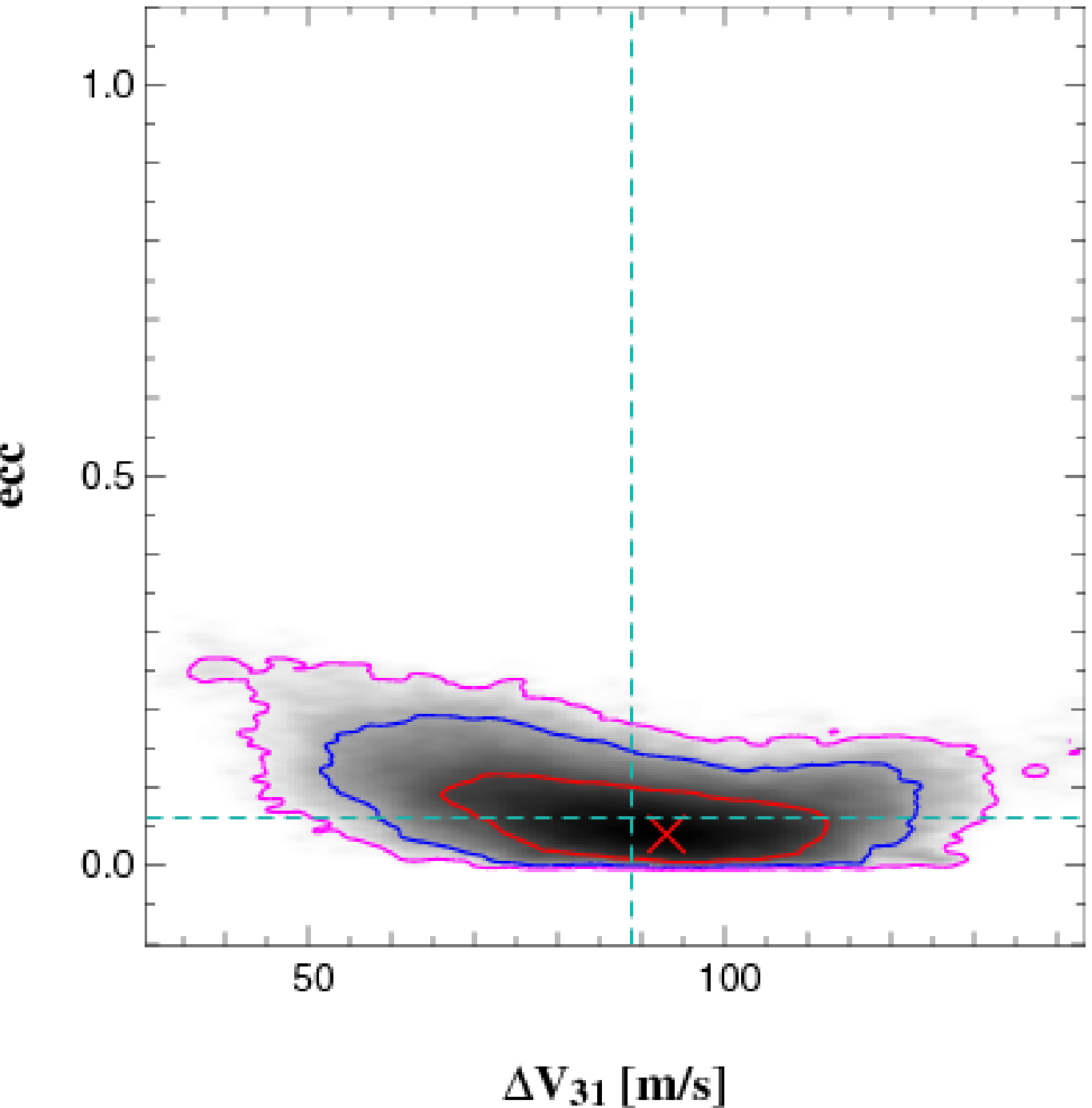}
   \includegraphics[width=6cm]{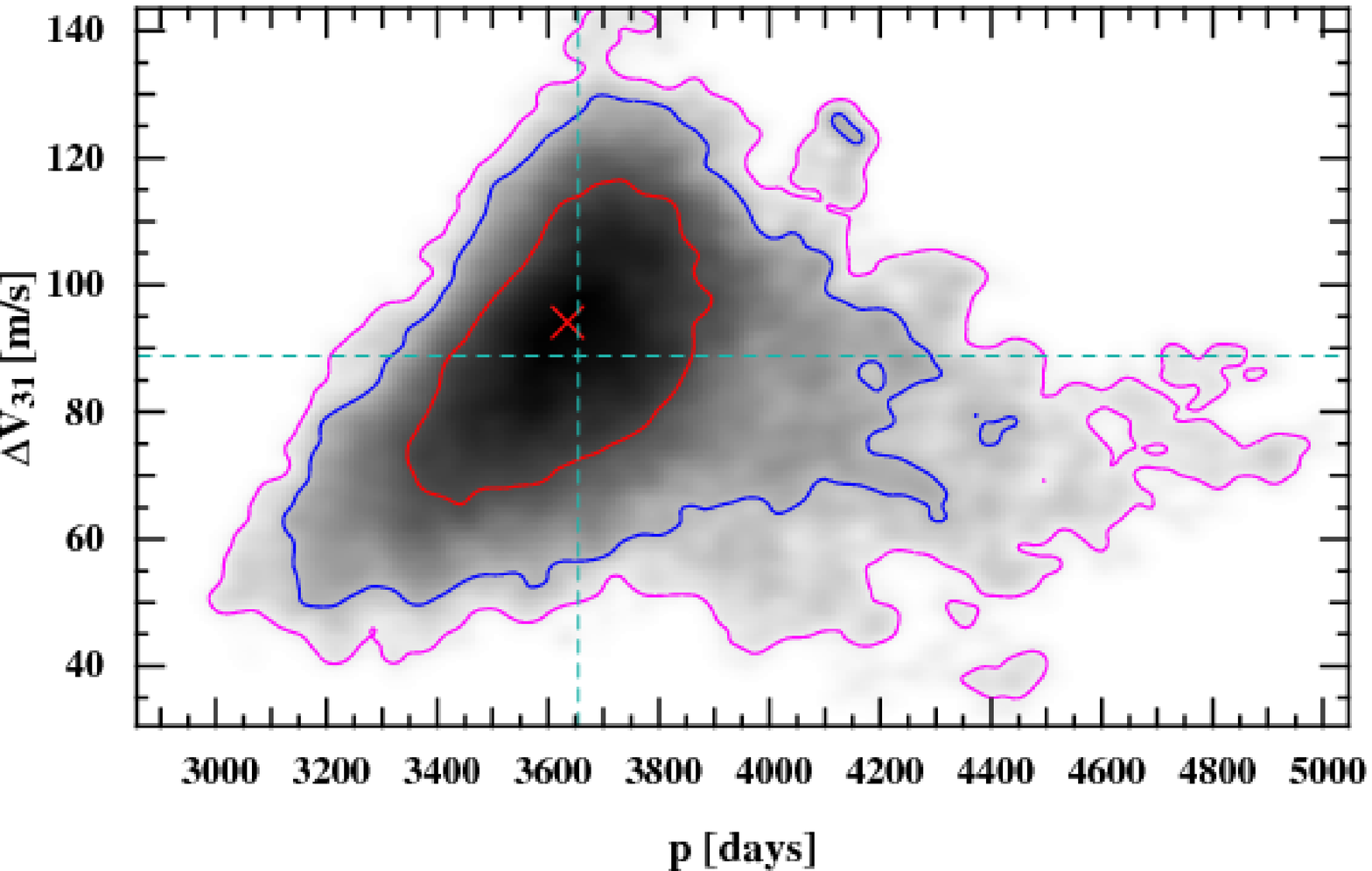}
    \caption{ Covariance between the semi-amplitude $K$ \textit{(left panels)}, the eccentricity $e$ \textit{(middle panels)}, the period $P$ \textit{(right panels)}, and the $\Delta$(RV)$_{\mathrm E-S}$ for the HD150706 \textit{(top panels)}, HD222155 \textit{(middle panels)} and HD24040 systems \textit{(bottom panels)}. The red, blue, and purple contour lines represent, respectively, the one, two, and three-$\sigma$ confidence intervals.}       
         \label{covariance}
   \end{figure*}

		Finally, we fitted using a Keplerian model the ELODIE and Keck measurements together with the corrected SOPHIE ones. The final orbital elements are listed in Table~\ref{param_p}. They were computed using 4.8~10$^6$ Monte Carlo simulations with a prior on the $\Delta$(RV)$_{\mathrm E-S}$ equals to the calibrated value and its uncertainty.  
		The uncertainties in the final parameters correspond to their 0.95 confidence intervals. The best-fit solution is consistent with a non-significant eccentric orbit,  $e$=0.38$^{+0.28}_{-0.32}$ that has a period of $P$=5894$^{+5584}_{-1498}$ days and a semi-amplitude $K$=31.1$^{+6.3}_{-4.8}$ms$^{-1}$. Taking into account the error bar in the stellar mass, HD150706b is a planet with a minimum mass $m_{p}\sin i$=2.71$^{+1.14}_{-0.66}$M$_{Jup}$ orbiting its star with a semi-major axis of 6.7$^{+4.0}_{-1.4}$ AU. In Fig.~\ref{fitHD150706}, the best-fit Keplerian model is superimposed on the ELODIE, Keck, and SOPHIE velocities. We also add plots in Fig.\ref{covariance} to illustrate the dependence of the K, P, and e parameters on $\Delta$(RV)$_{\mathrm E-S}$.
	
%p.14 du cahier: que reste-til dans les omc ?	
We did not find any indication of a second planet in the system with the current data set. From our solution, which has a dispersion of 15\,ms$^{-1}$, the RV residuals exclude an inner planet with $m_{\mathrm p}\sin i $ $>$ 1.3M$_{Jup}$. On the other hand, owing to the time span of 13.3~yr covered by our observations, we should not have missed an external planet that induces a drift larger than 1.1ms$^{-1}$yr$^{-1}$.   	
	
%--------------------------------------
\renewcommand{\arraystretch}{1.25}% 1.25 -> 140% en plus 
\begin{table}[h]
  \centering 
  \caption{Keplerian solution and inferred planetary parameters for HD150706b and HD222155b (see text for details).}
  \label{param_p}
\begin{tabular}{lcc}
\hline
\hline
Parameters   & HD150706b & HD222155b \\
\hline
$RV$$_{mean}$ \textsc{elodie}  [\,km\,s$^{-1}$] &   -17.2094$^{+0.0145}_{-0.0064}$   &  -43.9923$^{+0.0042}_{-0.0052}$   \\
$RV$$_{mean}$ \textsc{sophie}  [\,km\,s$^{-1}$] &   -17.1271$^{+0.0127}_{-0.0118}$  & -43.9007$^{+0.0129}_{-0.0114}$ \\ 
$RV$$_{mean}$ \textsc{keck}  [\,km\,s$^{-1}$] &   0.0322$^{+0.0074}_{-0.0075}$  &                             \\ 
$P$    [days]   &    5894$^{+5584}_{-1498}$  &   3999$^{+469}_{-541}$  \\
$K$          [\,m\,s$^{-1}$]     &   31.1$^{+6.3}_{-4.8}$  &   24.2$^{+6.4}_{-4.8}$  \\
$e$                & 0.38$^{+0.28}_{-0.32}$  &  0.16$^{+0.27}_{-0.22}$  \\
$\omega$    [deg]  &  132$^{+37}_{-33}$ &   137$^{+240}_{-52}$   \\
$T$$_{0}$    [JD]  &  58179$^{+4396}_{-1586}$    &  56319$^{+664}_{-498}$  \\
$m_{p}\sin i$  [M$_{\rm Jup}$] &    2.71$^{+1.14}_{-0.66}$~$^{\star}$ &  1.90$^{+0.67}_{-0.53}$~$^{\ast}$  \\
$a$   [AU]  &    6.7$^{+4.0}_{-1.4}$~$^{\star}$  &    5.1$^{+0.6}_{-0.7}$~$^{\ast}$  \\
$\sigma_{(O-C)}$   \textsc{elodie}    [\,m\,s$^{-1}$]  &   15.3   & 11.5  \\
$\sigma_{(O-C)}$   \textsc{sophie}    [\,m\,s$^{-1}$]  &   14.2   & 9.9  \\
$\sigma_{(O-C)}$   \textsc{keck}    [\,m\,s$^{-1}$]  &   6.2   &   \\
\hline
\end{tabular}
\begin{list}{}{}
\item[$^{\star}$] Assuming M$_{\star}$\,=\,1.17\,$\pm$\,0.12\,M$_{\odot}$ 
\item[$^{\ast}$] Assuming M$_{\star}$\,=\,1.13\,$\pm$\,0.11\,M$_{\odot}$
\end{list}

\end{table}
\renewcommand{\arraystretch}{1}% retour normal 
%---------------------------------------------------
	  
%p.15 du cahier barres d'erreur sur les masses et les AU	  

%-----------------------------------------------------------
%______________________________________________________________	

%-----------------------------------------------------------
%______________________________________________________________	

%----------------------------------------------------------------
 % Table 1 available electronically only
\onltab{1}{
 \begin{table*}
  \caption{Radial velocities of HD\,150706\ measured with ELODIE.} % DATA SUPER
  \label{table_rv}
\begin{tabular}{ccc}
\hline
\hline
BJD & RV & $\pm$$1\,\sigma$  \\
-2\,400\,000 & (km\,s$^{-1}$) & (km\,s$^{-1}$)   \\
\hline
50649.4745	& -17.194	& 0.009  \\
50942.5543	& -17.174	& 0.014  \\
50970.5200	& -17.196	& 0.012  \\
50970.5325	& -17.196	& 0.011  \\
50972.5962	& -17.197	& 0.010  \\
50973.4466	& -17.183	& 0.010  \\
50973.4591	& -17.178	& 0.012  \\
51269.6057	& -17.211	& 0.008  \\
51723.4471	& -17.189	& 0.009  \\
51728.4541	& -17.172	& 0.010  \\
51757.4065	& -17.189	& 0.008  \\
52081.4290	& -17.224	& 0.009  \\
52113.4120	& -17.255	& 0.008  \\
52137.3848	& -17.230	& 0.009  \\
52160.3491	& -17.198	& 0.013  \\
52163.3009	& -17.203	& 0.012  \\
52360.6618	& -17.243	& 0.012  \\
52361.6448	& -17.235	& 0.011  \\
52384.6408	& -17.227	& 0.013  \\
52388.6133	& -17.228	& 0.011  \\
52414.5642	& -17.250	& 0.009  \\
52451.4536	& -17.258	& 0.011  \\
52451.4667	& -17.244	& 0.012  \\
52482.3970	& -17.263	& 0.009  \\
52508.3361	& -17.243	& 0.010  \\
52747.6042	& -17.254	& 0.009  \\
52748.5678	& -17.247	& 0.009  \\
52751.5813	& -17.221	& 0.009  \\
52752.5889	& -17.224	& 0.008  \\
52773.5429	& -17.216	& 0.015  \\
52774.5403	& -17.228	& 0.018  \\
52776.5571	& -17.264	& 0.011  \\
52778.5717	& -17.245	& 0.008  \\
52797.4581	& -17.253	& 0.009  \\
52800.5201	& -17.238	& 0.009  \\
52801.4400	& -17.240	& 0.012  \\
52803.4532	& -17.253	& 0.009  \\
52806.4494	& -17.251	& 0.009  \\
52809.4272	& -17.243	& 0.010  \\
52812.4224	& -17.261	& 0.010  \\
52813.4814	& -17.262	& 0.009  \\
53160.4959	& -17.231	& 0.009  \\
53164.4548	& -17.236	& 0.010  \\
53223.3977	& -17.216	& 0.012  \\
53486.6043	& -17.214	& 0.011  \\
53490.5529	& -17.238	& 0.009  \\
53517.5451	& -17.237	& 0.009  \\
53900.4554	& -17.188	& 0.013  \\
\hline
\end{tabular}
\end{table*}
}% end of onltab
%---------------------------------------------------------------------------------

 %----------------------------------------------------------------
 % Table 2  available electronically only
\onltab{2}{
 \begin{table*}
  \caption{Radial velocities of HD\,150706\ measured with SOPHIE.}
  \label{table_rv2}
\begin{tabular}{ccc}
\hline
\hline
BJD & RV & $\pm$$1\,\sigma$  \\
-2\,400\,000 & (km\,s$^{-1}$) & (km\,s$^{-1}$)   \\    %HD150706_sophie3.1_selectedRV.rdb
\hline
54236.54555  &  -17.1770  & 0.0044  \\
54600.60889  &  -17.1808  & 0.0043  \\
54740.26854  &  -17.1806  & 0.0042  \\
54889.64197  &  -17.1265 &  0.0043  \\
55016.36337  &  -17.1479 &  0.0042  \\
55050.48029  &  -17.1448 &  0.0042  \\
55061.35488  &  -17.1510  & 0.0043  \\
55063.33252  &  -17.1943  & 0.0043  \\
55064.37769  &  -17.1777  & 0.0042  \\
55066.32264  &  -17.0975 &  0.0043  \\
55072.31036  &  -17.1709 &  0.0042  \\
55074.31585  &  -17.1952 &  0.0042  \\
55075.31406  &  -17.1521 &  0.0042  \\
55076.31876  &  -17.1233 &  0.0043  \\
55078.36177  &  -17.1370 &  0.0044  \\
55079.33895  &  -17.1174 &  0.0043  \\
55081.32237  &  -17.1624 &  0.0042  \\
55088.34489  &  -17.1409 &  0.0042  \\
55092.36418  &  -17.1885 &  0.0042  \\
55096.38831  &  -17.1455 &  0.0043  \\
55097.28451  &  -17.1610 &  0.0042  \\
55098.28223  &  -17.1467 &  0.0043  \\
55330.43771  &  -17.1678 &  0.0043  \\
55331.48420  &  -17.1692 &  0.0043  \\
55360.46481  &  -17.1541 &  0.0043  \\
55361.44615  &  -17.1640 &  0.0042  \\
55368.49765  &  -17.1355 &  0.0043  \\
55379.47954  &  -17.1746 &  0.0042  \\
55381.40204  &  -17.1568 &  0.0042  \\
55383.40329  &  -17.1307 &  0.0043  \\
55384.41523  &  -17.1591 &  0.0043  \\
55391.39659  &  -17.1614 &  0.0043  \\
55392.38517  &  -17.1462 &  0.0044  \\
55395.38753  &  -17.1201 &  0.0043  \\
55396.37419  &  -17.1606 &  0.0042  \\
55397.37673  &  -17.1841 &  0.0042  \\
55398.37019  &  -17.1750 &  0.0042  \\
55399.38494 &   -17.1430 &  0.0042  \\
55400.35673  &  -17.1436 &  0.0042  \\
55401.35616  &  -17.1372 &  0.0044  \\
55402.36160  &  -17.1653 &  0.0043  \\
55403.37580  &  -17.1580 &  0.0043  \\
55404.36871  &  -17.1336 &  0.0043  \\
55405.34437  &  -17.1079 &  0.0043  \\
55409.38627  &  -17.1421 &  0.0042  \\
55424.34173  &  -17.1670 &  0.0042  \\
55429.33811  &  -17.1585 &  0.0042  \\
55476.30322 &   -17.1748 &  0.0042  \\
55478.29437  &  -17.1674 &  0.0043  \\
55479.26619  &  -17.1527 &  0.0043  \\
55499.23547  &  -17.1677 &  0.0044  \\
55527.23017  &  -17.1695 &  0.0045 \\
55758.38710  &  -17.1495 &  0.0044 \\
\hline
\end{tabular}
\end{table*}
}% end of onltab
%---------------------------------------------------------------------------------

	\subsection{HD222155b, a Jupiter analog around a quiet star}
	We used the same methodology as for HD150706. First, we fixed the $\Delta$(RV)$_{\mathrm E-S}$ derived by the calibration (Appendix~\ref{RVES}) and used a Lomb Scargle periodogram to estimate the significance level of the detection of a long-period planet. With a fap$<$0.001, the highest peak corresponds to a period close to 4000 days.
	
	The ELODIE and SOPHIE RV data were then fitted with a Keplerian model. The eccentricity as well as the RV offset between the data sets were set as free parameters. The fitted offset -49$\pm$8ms$^{-1}$ agrees with the calibrated one within the error bars, -70$\pm$23ms$^{-1}$. The orbit has an insignificant eccentricity of $e$=0.26$\pm$0.24, a semi-amplitude of $K$=20.1\,ms$^{-1}$, and a period of 3259 days. The residuals to the fit $\sigma_{(O-C)}$=19.9ms$^{-1}$ are large compared to the mean error bar.
	
\begin{figure}
   \centering
   \includegraphics[width=8cm]{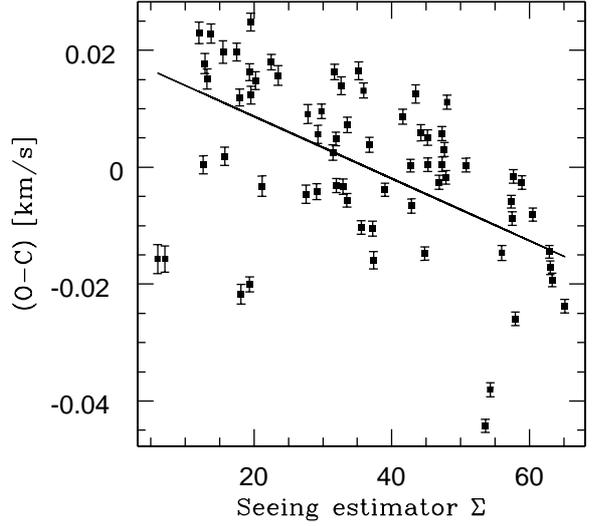}
      \caption{ SOPHIE residuals from the Keplerian fit of HD222155 as a function of the \textit{seeing estimator}. The best least squares linear fit is also plotted.   
                    }
         \label{SeeingOMCHD222155}
   \end{figure}

	The star is inactive and we do not expect any jitter as an astrophysical noise. On the other hand, the {\it seeing effect} is characterized in the SOPHIE data. In Fig.~\ref{SeeingOMCHD222155}, SOPHIE residuals are plotted as a function of the {\it seeing estimator} $\Sigma$. The correlation coefficient is equal to -0.51 and the Spearman coefficient to -0.5 with fap$<$10$^{-5}$, justifying a linear least squares fit to the data. 
	We corrected the SOPHIE RV for this trend $RV_{\mathrm corrected}$ [kms$^{-1}$]=$RV$ [kms$^{-1}$]+0.00068$\times \Sigma$. 	
%-----------------------------------------------------------
%______________________________________________________________	
	
	We then fitted a Keplerian model to the corrected SOPHIE RV and the ELODIE measurements.  The final orbital elements were computed based on 4.8~10$^6$ Monte Carlo simulations with a prior on the $\Delta$(RV)$_{\mathrm E-S}$ equals to the calibrated value and its uncertainty (and accounting for the correction on the SOPHIE RV). The  uncertainties correspond to the 0.95 confidence interval. They are listed in Table~\ref{param_p}. 
	The best-fit solution is an insignificant eccentric orbit ($e$=0.16$^{+0.27}_{-0.22}$) with a period $P$= 3999$^{+469}_{-541}$\,days and a semi-amplitude $K$=24.2$^{+6.4}_{-4.8}$~ms$^{-1}$.The corresponding planet has a minimum mass of $m_{p}\sin i$=1.90$^{+0.67}_{-0.53}$~M$_{Jup}$ and orbits HD222155 with a semi-major axis of 5.1$^{+0.6}_{-0.7}$~AU, taking into account the error bar in the stellar mass. In Fig.~\ref{fitHD222155}, the best-fit Keplerian model is superimposed to the ELODIE and SOPHIE velocities. {Plots in Fig.~\ref{covariance} show the covariance of the K, P, and e parameters with $\Delta$(RV)$_{\mathrm E-S}$.

%-----------------------------------------------------------
   \begin{figure}
   \centering
   \includegraphics[width=8cm]{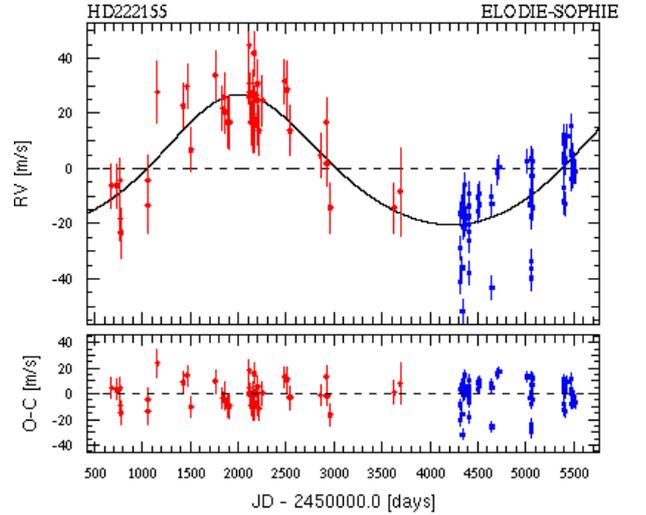}
      \caption{ ELODIE (red) and SOPHIE (blue) RV and residuals from the best-fit Keplerian model for HD222155 as a function of barycentric Julian date. The best-fit Keplerian model is represented by the black curve with a reduced $\chi^{2}$ equal to 2.2. The planet has a period of 10.9~yr in a non-significant eccentric orbit ($e$=0.16$^{+0.27}_{-0.22}$),  and a minimum mass of 1.90~M$_{Jup}$.               }
         \label{fitHD222155}
   \end{figure}
%______________________________________________________________	

%p.14 du cahier: que reste-til dans les omc ?	
No periodicity is detected in the RV residuals. The dispersion of the residuals, $\sigma$$_{(O-C)}$$\sim$11ms$^{-1}$, excludes an inner planet with $m_{p}\sin i $ $>$ 0.9M$_{Jup}$ and an external planet should not induce a drift larger than 0.8ms$^{-1}$yr$^{-1}$.

 %----------------------------------------------------------------
 % Table 3 available electronically only
\onltab{3}{
 \begin{table*}
  \caption{Radial velocities of HD\,222155\ measured with ELODIE.}   %HD222155_ELODIEsup.rdb
   \label{table_rv3}
\begin{tabular}{ccc}
\hline
\hline
BJD & RV & $\pm$$1\,\sigma$  \\
-2\,400\,000 & (km\,s$^{-1}$) & (km\,s$^{-1}$)   \\
\hline
50675.6213	& -44.001	& 0.008	  \\
50731.4575	& -44.001	& 0.008	  \\
50767.3732	& -44.013	& 0.008	  \\
50768.3859	& -43.999	& 0.008	  \\
50773.4144	& -44.018	& 0.009	  \\
51054.5803	& -43.999	& 0.009	  \\
51054.5931	& -44.008	& 0.010	  \\
51155.3039	& -43.967	& 0.011	  \\
51422.5284	& -43.972	& 0.008	  \\
51466.4467	& -43.965	& 0.008	  \\
51506.4018	& -43.988	& 0.008	  \\
51758.5948	& -43.961	& 0.009	  \\
51835.4570	& -43.973	& 0.008	  \\
51853.4017	& -43.969	& 0.009	  \\
51858.3551	& -43.974	& 0.008	  \\
51882.2985	& -43.978	& 0.009	  \\
51906.3044	& -43.978	& 0.010	  \\
52111.6150	& -43.950	& 0.008	  \\
52112.6131	& -43.964	& 0.007	  \\
52115.6158	& -43.968	& 0.008	  \\
52116.6066	& -43.969	& 0.008	  \\
52117.5774	& -43.967	& 0.007	  \\
52136.5939	& -43.978	& 0.008	  \\
52158.5965	& -43.967	& 0.008	  \\
52159.5882	& -43.980	& 0.009	  \\
52160.5985	& -43.977	& 0.011	  \\
52161.3818	& -43.975	& 0.010	  \\
52163.5326	& -43.977	& 0.012	  \\
52164.5088	& -43.953	& 0.008	  \\
52193.5460	& -43.968	& 0.008	  \\
52194.4816	& -43.979	& 0.007	  \\
52195.4841	& -43.970	& 0.007	  \\
52197.4701	& -43.964	& 0.009	  \\
52214.3879	& -43.981	& 0.009	  \\
52252.3722	& -43.970	& 0.009	  \\
52482.6168	& -43.963	& 0.008	  \\
52510.5121	& -43.966	& 0.010	  \\
52537.5497	& -43.981	& 0.009	  \\
52855.6104	& -43.990	& 0.008	  \\
52919.4593	& -43.978	& 0.009	  \\
52923.4549	& -43.993	& 0.008	  \\
52958.4170	& -44.009	& 0.009	  \\
53626.4855	& -44.009	& 0.009	  \\
53692.3277	& -44.003	& 0.016	  \\
\hline
\end{tabular}
\end{table*}
}% end of onltab
%---------------------------------------------------------------------------------

 %----------------------------------------------------------------
 % Table 4 available electronically only
\onltab{4}{
 \begin{table*}
  \caption{Radial velocities of HD\,222155\ measured with SOPHIE.}
  \label{table_rv4}
\begin{tabular}{ccc}
\hline
\hline
BJD & RV & $\pm$$1\,\sigma$  \\
-2\,400\,000 & (km\,s$^{-1}$) & (km\,s$^{-1}$)   \\
\hline
54311.59377	&  -43.9328	& 0.0043   \\
54313.60538	&  -43.9407	& 0.0042   \\
54314.54313	&  -43.9498	& 0.0044   \\
54316.55558	&  -43.9611	& 0.0042   \\
54335.58083	&  -43.9366	& 0.0042   \\
54337.58167	&  -43.9556	& 0.0042   \\
54338.52044	&  -43.9831	& 0.0042   \\
54344.49813	&  -43.9282	& 0.0043   \\
54345.47122	&  -43.9441	& 0.0042   \\
54346.50600	&  -43.9407	& 0.0042   \\
54349.57894	&  -43.9281	& 0.0042   \\
54358.54608	&  -43.9327	& 0.0042   \\
54359.50478	&  -43.9434	& 0.0042   \\
54360.50765	&  -43.9463	& 0.0042   \\
54375.43443	&  -43.9516	& 0.0042   \\
54404.38658	&  -43.9407	& 0.0043   \\
54405.27033	&  -43.9362	& 0.0042   \\
54406.32330	&  -43.9407	& 0.0041   \\
54408.27384	&  -43.9490	& 0.0042   \\
54408.28112	&  -43.9435	& 0.0042   \\
54409.26993	&  -43.9646	& 0.0042   \\
54409.27719	&  -43.9601	& 0.0042   \\
54496.23763	&  -43.9132	& 0.0044   \\
54497.24360	&  -43.9164	& 0.0043   \\
54499.25691	&  -43.9183	& 0.0044   \\
54514.26060	&  -43.9262	& 0.0043   \\
54516.25281	&  -43.9195	& 0.0043   \\
54644.60118	&  -43.9242	& 0.0043   \\
54645.60320	&  -43.9364	& 0.0042   \\
54646.59073	&  -43.9740	& 0.0042   \\
54702.59379	&  -43.9050	& 0.0043   \\
54724.51839	&  -43.9092	& 0.0042   \\
55010.60439	&  -43.9120	& 0.0042   \\
55050.49819	&  -43.9315	& 0.0042   \\
55051.55869	&  -43.9412	& 0.0042   \\
55054.62748	&  -43.9014	& 0.0045   \\
55055.63100	&  -43.9253	& 0.0042   \\
55056.54766	&  -43.9359	& 0.0042   \\
55057.55794	&  -43.9428	& 0.0043   \\
55058.53263	&  -43.9368	& 0.0046   \\
55059.52953	&  -43.9370	& 0.0043   \\
55061.54773	&  -43.9267	& 0.0042   \\
55062.53034	&  -43.9242	& 0.0042   \\
55063.56159	&  -43.9185	& 0.0042   \\
55066.54844	&  -43.9091	& 0.0043   \\
55072.48188	&  -43.9191	& 0.0043   \\
55073.54251	&  -43.9047	& 0.0042   \\
55391.58940	&  -43.9153	& 0.0042   \\
55392.59350	&  -43.9250	& 0.0043  \\
55393.60189	&  -43.9237	& 0.0042   \\
55398.60598	&  -43.9220	& 0.0042   \\
55400.61287	&  -43.9174	& 0.0042   \\
55401.60001	&  -43.8956	& 0.0043   \\
55402.58238	&  -43.9073	& 0.0042   \\
55404.58194	&  -43.9056	& 0.0043   \\
55405.51664	&  -43.9132	& 0.0042   \\
55409.52903	&  -43.9464	& 0.0043   \\
55433.57127	&  -43.9148	& 0.0042   \\
55476.42251	&  -43.9267	& 0.0042   \\
55478.43488	&  -43.9247	& 0.0042   \\
55479.40630	&  -43.9083	& 0.0043   \\
55483.45282	&  -43.9204	& 0.0042   \\
55484.35799	&  -43.9226	& 0.0042   \\
55495.44765	&  -43.9058	& 0.0044   \\
55498.31985	&  -43.9354	& 0.0041   \\
55505.38780	&  -43.9382	& 0.0042   \\
55523.40552	&  -43.9088	& 0.0043   \\

\hline
\end{tabular}
\end{table*}
}% end of onltab
%---------------------------------------------------------------------------------

\section{Refine the orbital parameters of previously announced long-period planets}

	The following targets were measured for the same subprogram and observed with the same strategy as that adopted for HD150706 and HD222155, which was detailed in Sect.~\ref{obs}.  

\subsection{HD24040b}
	
	Wright et al. (2007) presented the RV variability of HD24040b measured for this inactive G0V star using Keck data. At that time, the authors announced a companion with a period of between 10~yr and 100~yr and a minimum  mass in the range between 5\,M$_\mathrm{Jup}$ and 20\,M$_\mathrm{Jup}$. The stellar parameters can be found in Table~2 of Wright et al. (2007).
	
	 Our observations of HD24040, which were obtained during, for ELODIE September 1997 and December 2005, and for SOPHIE February 2008 and December 2010, have provided respectively 47 ELODIE and 21 SOPHIE measurements. The SOPHIE data with SNR$<$100 were removed (four observations) and we discarded three measurements for which there were abnormal flux level in the thorium-argon calibration lamp. 
	 
	 We combined both the ELODIE and SOPHIE datasets with the published Keck ones. We found that the best Keplerian fit converges with a RV offset between ELODIE and SOPHIE of $\Delta$(RV)$_{\mathrm E-S}$=-120$\pm$12\,ms$^{-1}$, which is significantly larger than the calibrated value of -74$\pm$23\,ms$^{-1}$ for this star with a $B-V$\,=\,$0.64$. Moreover, the RV diagram shows a clear trend, as seen in Fig.~\ref{HD24040}. We then fit the RV measurements with a Keplerian and a linear trend. 
	 We search for the \textit{seeing effect} in the residuals of the fit. The SOPHIE ($O-C$) data are plotted as a function of the \textit{seeing estimator} in Fig.~\ref{HD24040_see}. The correlation coefficient, which equals $-0.30$, is not significant with a 30\% probability that the two variables are uncorrelated. We removed from the study the measurement with the highest \textit{seeing estimator} value, which is certainly biased by the \textit{seeing effect}.  The final ELODIE and SOPHIE datasets are available electronically in Tables~\ref{table_rv5} and~\ref{table_rv6}. 
	 
 	  We re-adjusted the data with a simultaneous fit of a Keplerian and a linear trend. The RV offset is equal to -67$\pm$13\,ms$^{-1}$, in agreement within the error bars with the calibrated value. The final orbital elements are computed from 4.8~10$^6$ Monte Carlo simulations with a prior on the $\Delta$(RV)$_{\mathrm E-S}$ equals to the calibrated value and its uncertainty. Fig.~\ref{HD24040} shows the velocities as a function of time, as well as the fitted Keplerian orbit with a period of 3668 days and the linear trend of 3.85$^{+1.43}_{-1.29}$\,ms$^{-1}$yr$^{-1}$. The underlying linear drift easily explains why Wright et al. (2007) overestimated the period and the mass of HD24040b when fitting over a fraction of the orbital period. The solution is circular, $e$=0.04$^{+0.07}_{-0.06}$, with a semi-amplitude of $K$=47.4$^{+2.7}_{-2.6}$ms$^{-1}$. The inferred minimum mass of the companion, accounting for the uncertainty in the stellar mass, is 4.01$\pm$0.49~M$_{\mathrm Jup}$ with a semi-major axis of 4.92$\pm$0.38~AU (Table~\ref{param_p2}). 
	  The relations between the K, P, and e parameters and $\Delta$(RV)$_{\mathrm E-S}$ are plotted in Fig.~\ref{covariance}.
	  
	The residuals has a dispersion of  7.5\,ms$^{-1}$ and do not show any evidence of shorter period companions, and an inner planet with $m_{p}\sin i$\,$>$0.62\,M$_{\mathrm Jup}$ is excluded.

%-----------------------------------------------------------
   \begin{figure}
   \centering
   \includegraphics[width=8cm]{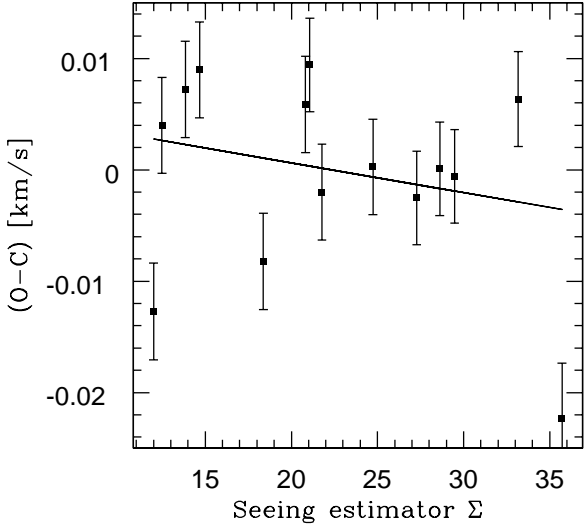}
      \caption{ HD24040 SOPHIE residuals from the Keplerian fit as a function of the \textit{seeing estimator}. The linear trend is insignificant. 
              }
         \label{HD24040_see}
   \end{figure}
%______________________________________________________________	

%-----------------------------------------------------------
   \begin{figure}
   \centering
   \includegraphics[width=8cm]{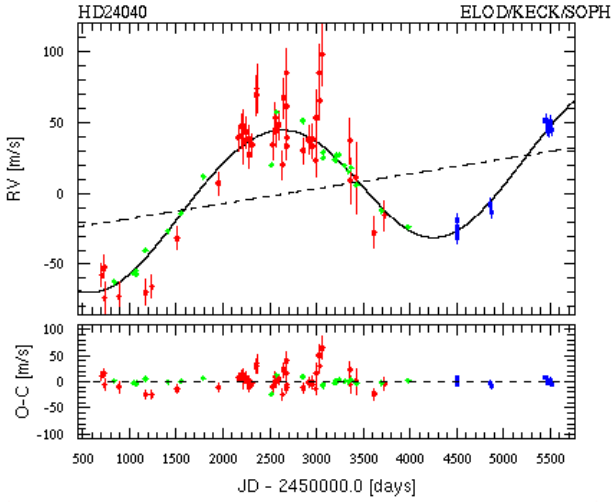}
      \caption{ ELODIE (red), Keck (green) SOPHIE (blue) RV and residuals to the best-fit Keplerian model (black curve) for HD24040 as a function of barycentric Julian date. It shows a 4.01~M$_{Jup}$ companion with an orbital period of 10.0~yr. A linear trend is fitted simultaneously pointing out the presence of a third body in the system.
              }
         \label{HD24040}
   \end{figure}
%______________________________________________________________	
	
%--------------------------------------
\renewcommand{\arraystretch}{1.25}% 1.25 -> 140% en plus 
\begin{table}[h]
  \centering 
  \caption{Keplerian solution and inferred planetary parameters for HD24040b with the combined measurements of ELODIE, SOPHIE, and the Keck RV data published by Wright et al. (2007).}
  \label{param_p2}
\begin{tabular}{lccc}
\hline
\hline
Parameters   &  HD24040b  \\
\hline
$RV$$_{mean}$ \textsc{elodie}  [\,km\,s$^{-1}$] &     -9.4003 $\pm$0.0040    \\
$RV$$_{mean}$ \textsc{sophie}  [\,km\,s$^{-1}$]   & -9.3118$^{+0.0150}_{-0.0156}$   \\ 
$RV$$_{mean}$ \textsc{keck}  [\,km\,s$^{-1}$] &     0.0311$^{+0.0039}_{-0.0036}$    \\

$RV$$_{linear}$ [\,m\,s$^{-1}$yr$^{-1}$]   &3.85$^{+1.43}_{-1.29}$  \\
$P$    [days]   &       3668$^{+169}_{-171}$    \\
$K$          [\,m\,s$^{-1}$]     &       47.4$^{+2.7}_{-2.6}$     \\
$e$                  &  0.04$^{+0.07}_{-0.06}$     \\
$\omega$    [deg]  &    154$^{+84}_{-54}$  \\
$T$$_{0}$    [JD]  &     54308$^{+859}_{-839}$    \\
$m_{p}\sin i$  [M$_{\rm Jup}$] &  4.01$\pm$0.49$^{1}$  \\
$a$   [AU]   &    4.92$\pm$0.38$^{1}$     \\
$\sigma_{(O-C)}$   \textsc{elodie}    [\,m\,s$^{-1}$]  &     14.9    \\
$\sigma_{(O-C)}$   \textsc{sophie}    [\,m\,s$^{-1}$]  &   7.3   \\
$\sigma_{(O-C)}$   \textsc{keck}    [\,m\,s$^{-1}$]  &     7.2   \\
\hline
\hline
\end{tabular}
\begin{list}{}{}
\item[$^{1}$] Assuming M$_{\star}$\,=\,1.18\,$\pm$\,0.10\,M$_{\odot}$
\end{list}
\end{table}
\renewcommand{\arraystretch}{1}% 1 -> 140% en plus 
%---------------------------------------------------

%----------------------------------------------------------------
 % Table 7 available electronically only
\onltab{5}{
 \begin{table*}
  \caption{Radial velocities of HD\,24040\ measured with ELODIE.}
  \label{table_rv5}
\begin{tabular}{ccc}
\hline
\hline
BJD & RV & $\pm$$1\,\sigma$  \\
-2\,400\,000 & (km\,s$^{-1}$) & (km\,s$^{-1}$)   \\
\hline	
50701.66050 & 	-9.458 & 	0.008	   \\
50730.60090 & 	-9.453 & 	0.007	   \\
50732.61430 & -9.452	 & 0.008	   \\
50736.61110	 & -9.474	 & 0.011	   \\  
50890.32770 & 	-9.473 & 	0.010	   \\
51174.38940 & 	-9.470 & 	0.009	   \\
51238.26700 & 	-9.466 & 	0.008	   \\
51510.56610 & 	-9.432	 & 0.008	   \\
51951.35990 & 	-9.393 & 	0.008	   \\
52164.65860 & 	-9.361 & 	0.009	   \\
52194.59920 & 	-9.360 & 	0.008	   \\
52199.59010 & 	-9.353 & 	0.009	   \\
52214.58190 & 	-9.353 & 	0.012	   \\
52218.58010 & 	-9.363 & 	0.009	   \\
52220.53190	 & -9.364	 & 0.015	   \\
52249.48790 & 	-9.357 & 	0.011	   \\
52278.40670	 & -9.362	 & 0.010	   \\
52280.38100	 & -9.373	 & 0.009	   \\
52308.36950 & 	-9.366	 & 0.008	   \\
52356.29560 & 	-9.331	 & 0.014	   \\
52361.29410 & 	-9.326	 & 0.017	   \\
52532.66570 & 	-9.366	 & 0.010	   \\
52559.61600 & 	-9.347	  & 0.011	   \\
52561.67100 & 	-9.357 & 	0.010	   \\
52565.62800 & 	-9.355	 & 0.010	   \\
52597.49410 & 	-9.352 & 	0.010   \\
52637.41740 & 	-9.380 & 	0.010	   \\
52647.40080 & 	-9.333 & 	0.014	   \\
52677.32400 & 	-9.339 & 	0.019	   \\
52678.36200 & 	-9.367 & 	0.009	   \\
52679.34660 & 	-9.315 & 	0.017	   \\
52681.38730 & 	-9.361 & 	0.010	   \\
52856.62940 & 	-9.370 & 	0.009	   \\
52919.55000 & 	-9.362 & 	0.010	   \\
52922.59580 & 	-9.363 & 	0.008	   \\
52954.56470 & 	-9.367 & 	0.008	   \\
52958.50350 & 	-9.362 & 	0.011	   \\
52993.44400 & 	-9.377 & 	0.011	   \\
52996.40330 & 	-9.347 & 	0.020	   \\
53030.34120 & 	-9.315 & 	0.020	   \\
53034.37680 & 	-9.335 & 	0.013	   \\
53059.28940	 & -9.302	 & 0.022	   \\
53358.41180 & 	-9.363	 & 0.016	   \\
53361.44080 & 	-9.391 & 	0.016	   \\
53421.30340 & 	-9.389 & 	0.025	   \\
53614.59350 & 	-9.428 & 	0.011	   \\
53726.40810 & 	-9.416 & 	0.011	   \\
\hline
\end{tabular}
\end{table*}
}% end of onltab
%---------------------------------------------------------------------------------
	
%----------------------------------------------------------------
 % Table 8 available electronically only
\onltab{6}{
 \begin{table*}
  \caption{Radial velocities of HD\,24040\ measured with SOPHIE.}
  \label{table_rv6}
\begin{tabular}{ccc}
\hline
\hline
BJD & RV & $\pm$$1\,\sigma$  \\
-2\,400\,000 & (km\,s$^{-1}$) & (km\,s$^{-1}$)   \\
\hline	
54503.29729 & 	-9.3414 & 	0.0043   \\
54504.26177 & 	-9.3422 & 	0.0043   \\
54505.26558 & 	-9.3482 & 	0.0043   \\
54506.27614 & 	-9.3425 & 	0.0043   \\
54507.30600 & 	-9.3503 & 	0.0043   \\
54857.33346 & 	-9.3258 & 	0.0043   \\
54872.32914 & 	-9.3290	 & 0.0043   \\
55448.63161 & 	-9.2676 & 	0.0042   \\
55476.65796	 &  -9.2763  &  0.0042  \\
55479.55853	 &  -9.2716 &    0.0043 \\ 
55483.55793	 &  -9.2753  &  0.0042 \\
55495.54768 & 	 -9.2659 &    0.0043 \\
55519.49131	 &  -9.2764  &    0.0042 \\
\hline
\end{tabular}
\end{table*}
}% end of onltab
%---------------------------------------------------------------------------------

\subsection{HD89307b}
	
	On the basis of observations acquired at the Lick Observatory since 1998, Fischer et al. (2009) published evidence of a companion of $m_{p}\sin i$\,=\,$1.78$\,$\pm$\,0.13\,M$_\mathrm{Jup}$ with a period of $2157$\,$\pm$\,$63$\,days and an eccentricity of $0.24~\pm~0.07$ in orbit around HD89307, which is a bright inactive G0 dwarf. We refer the reader to the Fischer et al. (2009) stellar parameters (cf. their Table 1). We note that their stellar values agree with those derived by Sousa et al. (2006) based on a SARG observation at TNG. 
	
	 We performed 46 ELODIE and 11 SOPHIE observations, respectively, between December 1997 and April 2006, and between December 2006 and February 2011. 
	 Our corresponding RV data are available electronically in Tables~\ref{table_rv7} and~\ref{table_rv8}. We combined our measurements with the Lick RV.  
	 Figure~\ref{HD89307} shows the Keplerian orbit and the residuals around the solution. The RV shift between ELODIE and SOPHIE, $\Delta$(RV)$_{\mathrm E-S}$=-66$\pm$12ms$^{-1}$, agrees with the calibrated one, -49$\pm$23ms$^{-1}$. No instrumental effect is observed in the SOPHIE data.
	 The planetary parameters agree with those of Fischer et al. (2009). The fitted parameters for the companion and their uncertainties corresponding to the 0.95 confidence interval computed from 5000 permutation simulations are listed in Table~\ref{param_p3}. Assuming a stellar mass of M$_{\star}$\,=\,1.03~M$_{\odot}$ and taking into account its uncertainty ($\pm$\,0.10\,M$_{\odot}$), we computed a planetary minimum mass of $m_{p}\sin i$\,=\,$2.0$\,$\pm$\,0.4M$_\mathrm{Jup}$ for the HD89307 companion, which is slightly higher than the previous published value. The planet has a little longer period of $2199$\,$\pm$\,$61$\,days and orbits at 3.34$\pm$0.17~AU. We confirm the probable eccentricity of the orbit with $e$=$0.25\,\pm\,0.09$. 
	 
	 The residuals do not show periodicity. With a dispersion of 8ms$^{-1}$, the residuals exclude the presence of an inner planet with a minimum  mass $m_{p}\sin i$\,$>$\,$0.5$\,M$_\mathrm{Jup}$.
	
%-----------------------------------------------------------
   \begin{figure}
   \centering
   \includegraphics[width=8cm]{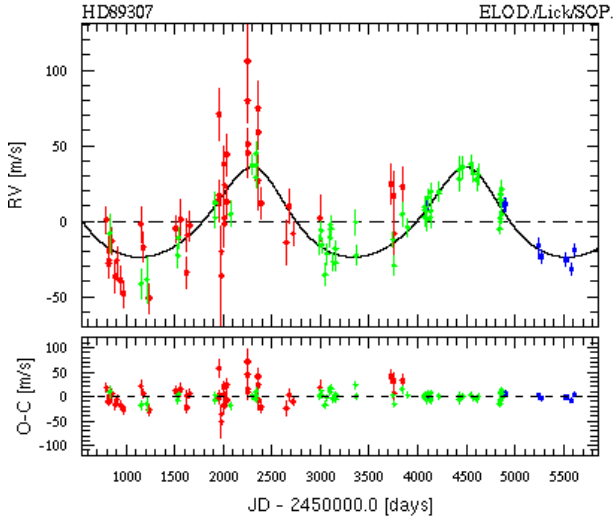}
      \caption{ELODIE (blue), Lick (green), and SOPHIE (red) RV and residuals from the best-fit Keplerian model (black curve) for HD89307 as a function of time. The fitted orbit corresponds to a planet with a minimum mass of 2.0~M$_{Jup}$, a period of 6.0~yr, and a slightly eccentricity orbit $e$=0.25$\pm$0.09.
              }
         \label{HD89307}
   \end{figure}
%______________________________________________________________	
	
 %----------------------------------------------------------------
 % Table 9 available electronically only
\onltab{7}{
 \begin{table*}
  \caption{Radial velocities of HD\,89307\ measured with ELODIE.}
  \label{table_rv7}
\begin{tabular}{ccc}
\hline
\hline
BJD & RV & $\pm$$1\,\sigma$  \\
-2\,400\,000 & (km\,s$^{-1}$) & (km\,s$^{-1}$)   \\
\hline
50796.70270 &	23.059  & 	0.009  \\
50822.59530 &	23.030  & 	0.009  \\
50823.59750 &	23.032 & 	0.009  \\
50824.59000 &	23.039 & 	0.009  \\
50859.52380 &	23.045 & 	0.009  \\
50888.48800 &	23.021 & 	0.010  \\
50888.50040 &	23.022 & 	0.010  \\
50908.39060 &	23.032 & 	0.009  \\
50942.37900 &	23.019 & 	0.009  \\
50972.34880 &	23.010 & 	0.009  \\
51153.72040 &	23.056 & 	0.012  \\
51177.62610 &	23.041 & 	0.010  \\
51234.51360 &	23.007 & 	0.010  \\
51507.69750 &	23.053 & 	0.009  \\
51562.57770 &	23.059 & 	0.013  \\
51623.44230 &	23.024 & 	0.010  \\
51626.41540 &	23.049 & 	0.011  \\
51655.37180 &	23.055 & 	0.010  \\
51953.65880 &	23.129 & 	0.017  \\
51954.57500 &	23.070 & 	0.012  \\
51955.52670 &	23.072 & 	0.011  \\
51956.49000 &	23.075 & 	0.010  \\
51978.48600 &	23.022 & 	0.033  \\
51979.46550 &	23.038 & 	0.011  \\
52007.35270 &	23.096 & 	0.012  \\
52008.38870 &	23.082 & 	0.013  \\
52009.47000 &	23.060 & 	0.019  \\
52010.43190 &	23.056 & 	0.013  \\
52033.36360 &	23.102 & 	0.014  \\
52038.38040 &	23.071 & 	0.010  \\
52248.71710 &	23.164 & 	0.025  \\
52248.72820 &	23.138 & 	0.015  \\
52252.68720 &	23.103 & 	0.016  \\
52252.70430 &	23.109 & 	0.011  \\
52356.44340 &	23.085 & 	0.019  \\
52356.45680 &	23.133 & 	0.018  \\
52358.43760 &	23.117 & 	0.011  \\
52388.41390 &	23.070 & 	0.010  \\
52649.68980 &	23.044 & 	0.015  \\
52677.55470 &	23.068 & 	0.011  \\
52719.43880 &	23.050 & 	0.008  \\
52996.58680 &	23.060 & 	0.015  \\
53726.63150 &	23.083 & 	0.013  \\
53750.72660 &	23.075 & 	0.016  \\
53756.57870 &	23.050 & 	0.014  \\
53842.36680 &	23.081 & 	0.013  \\
\hline
\end{tabular}
\end{table*}
}% end of onltab
%---------------------------------------------------------------------------------
	
 %----------------------------------------------------------------
 % Table 10 available electronically only
\onltab{8}{
 \begin{table*}
  \caption{Radial velocities of HD\,89307\ measured with SOPHIE.}
  \label{table_rv8}
\begin{tabular}{ccc}
\hline
\hline
BJD & RV & $\pm$$1\,\sigma$  \\
-2\,400\,000 & (km\,s$^{-1}$) & (km\,s$^{-1}$)   \\
\hline
54097.67291 &	23.1332 & 	0.0043  \\
54097.68020 &	23.1324 & 	0.0043  \\
54889.51926 &	23.1348 & 	0.0043  \\
54904.50409 &	23.1351 & 	0.0045  \\
55238.50655 &	23.1079	 & 0.0045  \\
55267.54064 &	23.1007 & 	0.0046  \\
55525.68715 &	 23.0984 &   0.0043 \\ 
 55526.70981 &	 23.0989  & 0.0044 \\ 
 55527.67595 &	 23.0979  & 0.0042  \\
 55586.56773 &	 23.0925  &   0.0043\\
 55615.48424 &	 23.1047  & 0.0044  \\
\hline
\end{tabular}
\end{table*}
}% end of onltab
%---------------------------------------------------------------------------------

	\subsection{HD154345b}
	Wright et al. (2008) reported the detection of HD154345b with a minimum mass of $m_{p}$$\sin$$i\,=\,0.94\,\pm\,0.09$\,M$_\mathrm{Jup}$,  an orbital period P\,=\,3539\,$\pm$\,66\,d, and an insignificant eccentricity of $e=0.044\pm0.046$. The host star is a bright quiet G8V (m$_{V}$\,=\,$6.7$) star with an estimated mass $M_{\star}$\,=\,$0.88$\,$\pm$\,$0.09$\,M$_{\odot}$. The stellar parameters can be found in Table 1 of Wright et al. (2008). 
	
	The star was also observed by ELODIE and SOPHIE with, respectively, 49 and 15 measurements spanning 12.2~yr and 3.2~yr. Three measurements were removed from the SOPHIE sample owing to the abnormal flux levels of the thorium-argon calibration lamp during observations and one because it was of too low SNR. We combined these measurements with the Keck RV and fit them with a Keplerian model. The best-fit solution converges with a RV offset between ELODIE and SOPHIE of $\Delta$(RV)$_{\mathrm E-S}$=107$\pm$6ms$^{-1}$, in agreement with the calibrated one of 108$\pm$23ms$^{-1}$ for this star with a $B-V$=0.73 (see App.~\ref{RVES}). 
	We searched for any \textit{seeing effect} in the SOPHIE data. As in HD24040, we found that only one measurements was significantly affected by the instrumental effect. We removed this data point from the sample and fit the three data sets with a Keplerian model. The final ELODIE and SOPHIE datasets are available electronically in Tables~\ref{table_rv9} and~\ref{table_rv10}.
	We found that the best-fit solution has an equivalent mass and period to Wright et al. (2008) values and an insignificant eccentricity $e$=0.26$\pm$0.15. The fitted parameters for the companions and their uncertainties computed from 5000 permutations simulations and their 0.95 confidence intervals are listed in Table~\ref{param_p3}. The final RV offset is equal to $\Delta$(RV)$_{\mathrm E-S}$=112$\pm$10\,ms$^{-1}$. We inferred a minimum mass of 1.0$\pm$0.3~M$_{\mathrm Jup}$, and semi-major axis of 4.3$\pm$0.4~AU. The error bars take into account the uncertainty in the stellar mass. The best-fit solution is plotted in Fig.~\ref{HD154345}. No significant variability is found in the residuals, and for a total dispersion of 4\,ms$^{-1}$, an inner planet with $m_{p}\sin i$$>$0.3\,M$_{\mathrm Jup}$ is not allowed.
	
%-----------------------------------------------------------
   \begin{figure}
   \centering
    \includegraphics[width=8cm]{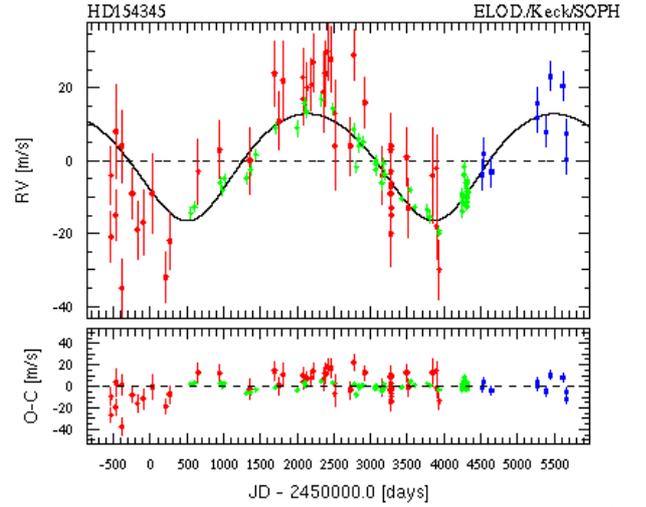}
      \caption{ ELODIE (red), Keck (green), and SOPHIE (blue) RV and residuals of the best-fit Keplerian model (black curve) for HD154345 as a function of barycentric Julian date. The companion has a period of 9.7~yr, and a minimum mass of 1.0~M$_{Jup}$.
              }
         \label{HD154345}
   \end{figure}
%______________________________________________________________	

%--------------------------------------
\begin{table}[h]
  \centering 
  \caption{Keplerian solution and inferred planetary parameters for HD154345b and HD89307b with the combined measurements of ELODIE, SOPHIE, and the already published RV data.}
  \label{param_p3}
\begin{tabular}{lccc}
\hline
\hline
Parameters   & HD154345b &  HD89307b \\
\hline
$RV$$_{mean}$ \textsc{elodie}  [\,km\,s$^{-1}$] &   -46.954 $\pm$ 0.005   &    23.067 $\pm$0.005   \\
$RV$$_{mean}$ \textsc{sophie}  [\,km\,s$^{-1}$] &   -46.842 $\pm$ 0.005  & 23.133 $\pm$0.007 \\ 
$RV$$_{mean}$ \textsc{original RV}  [\,km\,s$^{-1}$] &   -0.004 $\pm$ 0.005$^{A}$   &    0.007 $\pm$0.005$^{B}$  \\

$P$    [days]   &    3538$\pm$300  &   2199$\pm$61 \\
$K$          [\,m\,s$^{-1}$]     &  17.0 $\pm$ 3.7  &      32.4$\pm$4.5  \\
$e$                & 0.26 $\pm$0.15  &    0.25$\pm$0.09 \\
$\omega$    [deg]  &  -14$\pm$160 &  14 $\pm$  33 \\
$T$$_{0}$    [JD]  &  54701 $\pm$ 1000    &  54549$\pm$190\\
$m_{p}\sin i$  [M$_{\rm Jup}$] &    1.0 $\pm$0.3$^{1}$ &     2.0$\pm$0.4$^{2}$\\
$a$   [AU]  &    4.3$\pm$0.4$^{1}$  &     3.34$\pm$0.17$^{2}$\\
$\sigma_{(O-C)}$   \textsc{elodie}    [\,m\,s$^{-1}$]  &   12.9   &    19.5 \\
$\sigma_{(O-C)}$   \textsc{sophie}    [\,m\,s$^{-1}$]  &   6.6   & 4.1 \\
$\sigma_{(O-C)}$   \textsc{original RV}    [\,m\,s$^{-1}$]  &   2.9$^{A}$   &   8.4$^{B}$ \\

\hline
\end{tabular}
\begin{list}{}{}
\item[$^{A}$] Keck RV (Wright et al. 2008)
\item[$^{B}$] Lick RV (Fischer et al. 2009)
\item[$^{1}$] Assuming M$_{\star}$\,=\,0.88\,$\pm$\,0.09\,M$_{\odot}$
\item[$^{2}$] Assuming M$_{\star}$\,=\,1.03\,$\pm$\,0.10\,M$_{\odot}$

\end{list}

\end{table}

%---------------------------------------------------

 %----------------------------------------------------------------
 % Table 5 available electronically only
\onltab{9}{
 \begin{table*}
  \caption{Radial velocities of HD\,154345\ measured with ELODIE.}
  \label{table_rv9}
\begin{tabular}{ccc}
\hline
\hline
BJD & RV & $\pm$$1\,\sigma$  \\
-2\,400\,000 & (km\,s$^{-1}$) & (km\,s$^{-1}$)   \\
\hline
49464.62350 &	-46.955 & 	0.008\\	
49465.61330 &	-46.972 & 	0.007\\	
49527.49660 &	-46.966 & 	0.007\\	
49530.57980 &	-46.943 & 	0.013\\	
49611.28600 &	-46.986 & 	0.008\\	
49612.33000 &	-46.947 & 	0.010\\	
49753.70470 &	-46.960 & 	0.007\\	
49824.57510 &	-46.970 & 	0.008\\	
49905.46200 &	-46.968 & 	0.009\\	
50026.25970 &	-46.960 & 	0.011\\	
50210.55670 &	-46.983 & 	0.007\\	
50265.48900 &	-46.973 & 	0.008\\	
50651.40100 &	-46.954 & 	0.009\\	
50943.59820 &	-46.948 & 	0.008\\	
51357.44270 &	-46.951 & 	0.009\\	
51693.49520 &	-46.927 & 	0.009\\	
51755.38700 &	-46.940 & 	0.008\\	
51805.33550 &	-46.929 & 	0.011\\	
52080.46120 &	-46.928 & 	0.008\\	
52081.45600 &	-46.934 & 	0.007\\	
52136.38430 &	-46.931 & 	0.007\\	
52195.27040 &	-46.930 & 	0.007\\	
52217.23170 &	-46.924 & 	0.008\\	
52357.64990 &	-46.932 & 	0.011\\	
52385.56940 &	-46.927 & 	0.009\\	
52415.44280 &	-46.921 & 	0.008\\	
52459.46540 &	-46.923 & 	0.009\\	
52513.39240 &	-46.947 & 	0.012\\	
52513.40130 &	-46.938 & 	0.009\\	
52723.67010 &	-46.947 & 	0.008\\	
52772.62020 &	-46.922 & 	0.007\\	
52922.27230 &	-46.935 & 	0.007\\	
53160.51780 &	-46.955 & 	0.008\\	
53275.27350 &	-46.948 & 	0.008\\	
53275.35870 &	-46.960 & 	0.010\\	
53276.27840 &	-46.954 & 	0.009\\	
53277.33430 &	-46.958 & 	0.008\\	
53278.26730 &	-46.947 & 	0.009\\	
53278.36430 &	-46.971 & 	0.009\\	
53280.26290 &	-46.964 & 	0.008\\	
53280.33920 &	-46.957 & 	0.009\\	
53281.25430 &	-46.954 & 	0.008\\	
53281.33470 &	-46.966 & 	0.008\\	
53490.57280 &	-46.950 & 	0.008\\	
53516.60050 &	-46.964 & 	0.008\\	
53845.61620 &	-46.955 & 	0.013\\	
53898.53400 &	-46.969 & 	0.009\\	
53900.47320 &	-46.953 & 	0.009\\	
53935.45520 &	-46.981 & 	0.008\\
\hline
\end{tabular}
\end{table*}
}% end of onltab
%---------------------------------------------------------------------------------

%----------------------------------------------------------------
 % Table 6 available electronically only
\onltab{10}{
 \begin{table*}
  \caption{Radial velocities of HD\,154345\ measured with SOPHIE.}
  \label{table_rv10}
\begin{tabular}{ccc}
\hline
\hline
BJD & RV & $\pm$$1\,\sigma$  \\
-2\,400\,000 & (km\,s$^{-1}$) & (km\,s$^{-1}$)   \\
\hline
54515.70527  & 	-46.8432	 & 0.0042   \\
54539.69224 & 	-46.8375 & 	0.0042   \\
54647.44384 & -46.8425	 & 0.0043   \\
55268.65075	 & -46.8275 & 	0.0041   \\
55272.64768	 & -46.8235	 & 0.0041   \\
55391.38714	 & -46.8314	 & 0.0042   \\
55448.30372   &   -46.8162  &  0.0042	   \\	 
55619.69751    &  -46.8187  &   0.0042	 \\    	
55671.61674   &   -46.8318   &   0.0042	     \\	
55671.62197  &    -46.8391   &   0.0042	     \\	
\hline
\end{tabular}
\end{table*}
}% end of onltab
%---------------------------------------------------------------------------------

\section{Are we observing magnetic cycles ?} %p.16 du cahier
		
		It is only recently that the discoveries of planets with orbital periods reaching the range where stellar magnetic cycles have been observed (from 2.5 to 25 years, Baliunas et al. 1995), have been achievable.
		 A magnetic cycle could induce RV variations with the periodic modification of the number of spots and plages on the stellar photosphere (as observed on the Sun on a 11-year period), related to changes in the convection pattern and/or other mechanisms such as meridional flows (Beckers 2007, Makarov 2010), owing to the magnetic field created by dynamo. 
		The $\log$R'$_{\mathrm HK}$ index computed from the \ion{Ca}{II}~H\&K lines is sensitive to the presence of plages in the stellar chromosphere and is a reliable means of monitoring the magnetic cycle.
		  
		 Dedicated RV observations of stars with known magnetic cycles (Santos et al. 2010a, Gomes da Silva et al. 2012) have measured weak correlations between active lines indices (\ion{Ca}{II}~H\&K, H$\alpha$, \ion{Na}{I}) and RV, as well as in the parameters of the CCF. However, these studies have been limited by a narrow range of spectral types, respectively, early-K and early-M dwarfs. 
		 
		 On the other hand, high-precision stabilized fiber-fed spectrographs that observe in the visible such as HARPS or SOPHIE can accurately measure the flux in the \ion{Ca}{II}~H\&K lines. They can monitor with high precision the variation with time in the $\log$R'$_{\mathrm HK}$ index (Lovis et al. 2011b). While searching for planets, HARPS RV measurements have revealed stellar magnetic cycles (Moutou et al. 2011, S\'egransan et al. 2011, Dumusque et al. 2011). Lovis et al. (2011b) used the HARPS sample to identify activity cycles and derive relations between the RV and CCF parameter variations as a function of the $R'_{HK}$ variability. These relations depend on the stellar effective temperature and could be used to estimate the RV jitter produced by a magnetic cycle.  
		  
		The FWHM or contrast of the CCF are insufficiently accurate in the ELODIE or SOPHIE measurements to permit us to examine their variations. In addition, the accuracy of the ELODIE BIS is too low to be sensitive to the effect of a magnetic cycle. 
		 Moreover, the use of the thorium-argon lamp during the observations leads to polluted light on the CCD detector that prevents the measurement of the flux inside the active lines for ELODIE spectra. Only SOPHIE measurements of active lines can be used on a shorter timescale ($\sim$3 years) to check for stellar variability. Our observations alone cannot provide any conclusions about the existence of magnetic cycles on the reported stars. Pursuing further observations is therefore needed. 
		 
		Nevertheless, we measured the Pearson and Spearman correlation coefficients between the $\log$R'$_{\mathrm HK}$ and the RV values extracted from the SOPHIE data. For the only active star of the sample, HD150706, we averaged the measurements into bins of 30 days to remove the effect of the rotational period. We tested the significance of these coefficients with 100,000 Monte Carlo simulations of shuffled data. We did not find any correlation that could place in doubt the planetary hypothesis. 				
		
		We can assessed the planetary hypothesis using the results of Lovis et al. (2011b). We observed that in their Fig.~19 the maximal RV amplitude induced by a magnetic cycle is 12\,ms$^{-1}$. The detected RV semi-amplitudes reported in our paper are all greater than 17\,ms$^{-1}$, the smallest one being measured for HD154345. Using Eq.~9 of Lovis et al. (2011b) and the calcium index variations published by Wright et al. (2008), we calculated that the expected RV semi-amplitude due to an active cycle for HD154345 is 3.65$\pm$0.41\,ms$^{-1}$, which is far below the observed one. We also found that the $\log$R'$_{\mathrm HK}$ semi-amplitude needed to induce the RV variation measured in HD89307 is two times higher than the highest modulation observed by Lovis et al. (2011b) owing to magnetic cycle (cf. their Fig.~10). We concluded that the most likely explanation of the observed RV variations for our stars is the planetary hypothesis.

\section{Concluding remarks}

We have presented the detection of two new Jupiter-like planet candidates around HD150706 and HD222155 with combined measurements from the ELODIE and the SOPHIE spectrographs, which were mounted successively on the 1.93-m telescope at the OHP.
Orbiting farther than 5~AU from their parent stars, the planets have minimum masses of 2.71~M$_{Jup}$ and 1.90~M$_{Jup}$, respectively.  We have also published the first reliable orbit for HD24040b, which is another gaseous long-period planet. We determined a minimum mass of 4.01~M$_{\mathrm Jup}$ for this planet in a 10.0~yr orbit at 4.92~AU. We have presented evidence of a third companion in this system. Moreover, we have refined the planetary parameters of two others Jupiter-analogs, HD154345b and HD89307b, by combining our RV data with, respectively, the Keck and the Lick observatories measurements. We obtain parameter values in agreement with those of Wright et al. (2008) and Fischer et al. (2009). \\

HD150706 is an active star and the signature of its effect was detected in the BIS of the CCF. We corrected the SOPHIE measurements for the jitter effect. The four other stars are quiet with $\log$R$'_\mathrm{HK}$ values lower than $-4.9$.  In contrast, the SOPHIE measurements are affected by instrumental uncertainties caused by seeing variations, which we partly corrected. 

The amplitudes of the RV variations are greater in the case of all stars than for all the reported active cycles in the literature (Baliunas et al. 1995, Lovis et al. 2011b). We did not find any long-term correlations between the RV and the activity index in the SOPHIE measurements. We concluded that the most likely explanation of the observed RV variations is the presence of a planet. \\

 In IRAC and MIPS data acquired by \textit{Spitzer}, Meyer et al. (2004) detected for HD150706, an infrared excess at 70~$\mu$m, an upper limit at 160~$\mu$m, and no evidence of an excess at $\lambda$$<$35~$\mu$m. They interpreted their observations as evidence of a dust disk surrounding the star with a hole devoid of dust that has an inner radius of at least 20~AU. The authors proposed that the presence of an exoplanet could explain the inner edge of the outer dust disk. The SOPHIE and ELODIE RV data sets show evidence of a large companion at less than 20~AU around HD150706. With a minimum  mass of 2.71~M$_{Jup}$, HD150706b orbiting at 6.7$^{+4.0}_{-1.4}$~AU may keep clear the inner region of the disk. \\

     Examining the current distribution of the exoplanet candidate periodicities discovered by RV (Fig.~\ref{all}), we observe a drop after $\approx$~4~AU. These long-period planets are part of a new parameter space, which have been achieved thanks to the extension of the timelines of RV surveys to longer than 15 years. The current paper increases to nineteen the number of planets further than 4~AU characterized by the RV measurements (Table~\ref{liste}). With partial observations (i.e. where the orbital period was not completely covered) and a small number of objects, it has been difficult to establish significant statistical trends. 
   
     Nevertheless, in Fig.~\ref{zoom}, we focused on the planets discovered beyond 4~AU. We remark that no very massive planet ($>$8M$_{\mathrm Jup}$) was found beyond 4~AU, in spite of a RV bias detection toward high-mass objects.  We emphasize that the only one, HD106270b, is a particular object reported by Johnson et al. (2011) as a very massive planet ($m_{p} \sin i$\,=\,11\,M$_{\mathrm Jup}$) orbiting a subgiant.  The occurrence rate of planets with minimum masses higher than 8~M$_{\mathrm Jup}$ is $1/19$ for semi-major axes a$>$4~AU compared to $27/196$ ($\approx1/7$) for smaller orbits with $1<a<4$AU\footnote{Statistics were derived from the catalog of the website \textit{exoplanet.eu}}. Assuming a binomial distribution, this implies that 13.8$\pm$2.5\% of the planets with semi-major axes in the range $1<a<4$AU  and 5.3$\pm$22.3\% for those with semi-major axes a$>$4~AU have minimum masses higher than 8M$_{\mathrm Jup}$. The last error bar illustrates the effects of small number statistics. It is unlikely that these host stars would have been discarded from planet surveys as single-lined spectroscopic binaries: for instance, a 8\,M$_{\mathrm Jup}$ orbiting in 4000\,days a one solar-mass star induces a RV semi-amplitude of 102\,ms$^{-1}$ for a circular orbit, which leads to a typical linear slope of $\sim$\,37\,ms$^{-1}$yr$^{-1}$. If this result is not caused by an observational selection effect, and if we assume that these objects are formed by core-accretion, an explanation could be that these planets did not migrate a lot, preventing a large accumulation of material. Or else, the disk could dissipate when these planets formed preventing them from migrating and growing in mass. We remark that Mordasini et al. (2012) highlighted that a decrease in
frequency of giant planets at larger distance ($>\sim$5AU) is a solid prediction of the core accretion theory. If this absence of very massive planets beyond 4-5 AU becomes statistically significant, it is an important result for formation theory.
     
      If we consider only the planets published with a complete coverage of their orbits, they are mostly non-eccentric ($e$$<$0.25). In contrast, those with incomplete coverages are almost entirely eccentric ($e$$>$0.25), reflecting that eccentric orbits are more easily detected for periods longer that the observation times (Cumming 2004). However, the eccentricity distribution of these planets agrees with the current observation of a significant dispersion in eccentricities. But we emphasize that a slight eccentricity may hide a longer period planet.
    
     We observe that these planets are found instead in multiplanetary systems (10 of 19 candidates). This could be due to an observational bias as systems with planets are preferentially followed-up. The multiple systems (including HD24040) are plotted in Fig.~\ref{multi}. We remark that the sample includes two of the most populated systems known, $\mu$~Ara (HD160691) and 55~Cnc (HD75732) with respectively, four and five planets. For these systems, the longer period planet is the most massive one. HD134987 and HD183263 have similar configurations that our Jupiter-Saturn system with a lower mass planet outside. Two stars, HD187123 and HD217107, also host a short-period giant planet.
     
   Most of the host stars are G-type dwarf stars. This is clearly an observational bias, as G-type stars were the first spectral type to be targeted by RV surveys. Fig.~\ref{metal} shows the distribution of the host star metallicity. These detections come from different surveys and samples, and it is not easy to compare the occurrence rates. Nevertheless, a first observation would be that giant gaseous planets appear to occur significantly around stars that are more metal-rich than average (Santos et al. 2004, Fischer \& Valenti 2005). 
      
    These giants planets are supposed to be formed beyond the "snow line". According to the models of planet formation and orbital evolution, giant planets migrate inward on a timescale comparable with the lifetime of the protoplanetary disk. These giants planets with long-orbital periods should have neither migrated or they have followed a scenario that brings them to this location. They may have formed at the same time as the disk dissipated preventing them from migrating. They also may have interacted with other planets in the system causing them to migrate outwards or hamper their migration. For example, inward migration could be avoided by resonance trapping if the mass of the outer planet is a fraction of the mass of the inner planet, as in the Jupiter-Saturn case (Masset \& Snellgrove, 2001, Morbidelli \& Crida, 2007).\\

 %-----------------------------------------------------------
   \begin{figure}
   \centering
   \includegraphics[width=9cm]{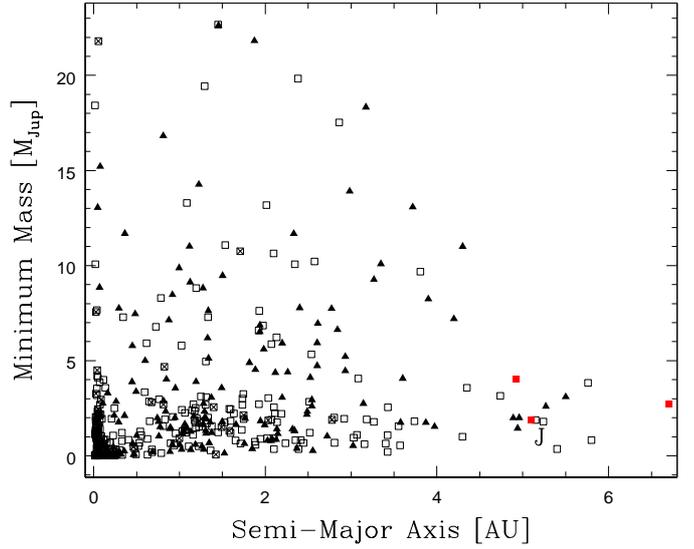}
      \caption{Minimum mass as a function of the semi-major axis for all planets detected by RV and transit surveys. Empty squared symbols (filled triangles) represent planets with eccentricities lower (higher) than 0.25. Crosses indicate fixed eccentricities at $e$=0. Jupiter is on the plot. Red points are the Jupiter-like planets characterized in this paper: HD150706b, HD222155b, and HD24040b.   
              }
         \label{all}
   \end{figure}
%______________________________________________________________	
	 
  %-----------------------------------------------------------
   \begin{figure}
   \centering
   \includegraphics[width=9cm]{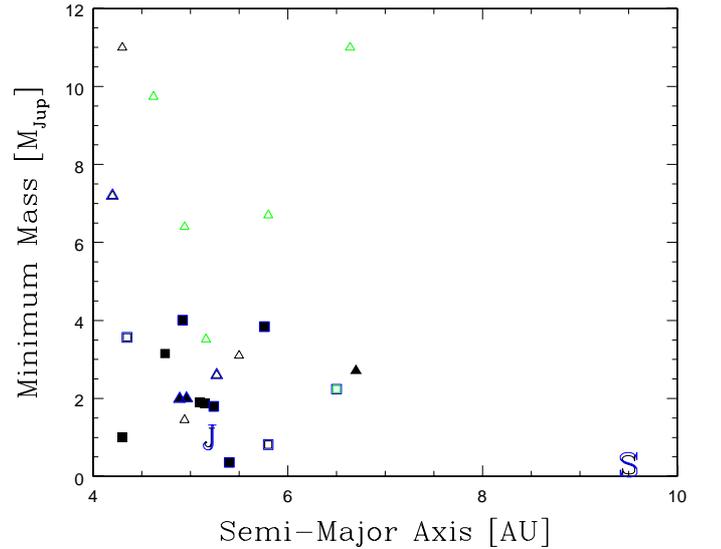}
      \caption{Minimum mass as a function of the semi-major axis for planets detected by RV with $a>$4~AU. Empty markers shows the incomplete orbits, while filled ones represent complete orbits. Squares and triangles represent, respectively, for low ($e$$<$0.25) and high ($e$$>$0.25) eccentricity orbits. The markers surrounded by blue show multiple systems. The green points indicate the higher mass planets announced by Marmier et al. (in prep., private communication).
              }
         \label{zoom}
   \end{figure}
%______________________________________________________________	

   %-----------------------------------------------------------
   \begin{figure}
   \centering
   \includegraphics[width=8.5cm]{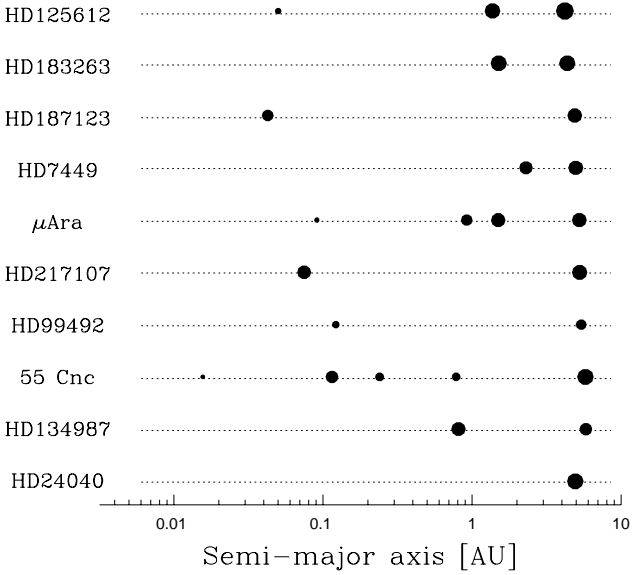}
      \caption{Multiple systems with semi-major axis greater than 4~AU. The size of the dots shows the minimum mass of the planet on a $\log$ scale.
              }
         \label{multi}
   \end{figure}
%______________________________________________________________	

  %-----------------------------------------------------------
   \begin{figure}
   \centering
   \includegraphics[width=9cm]{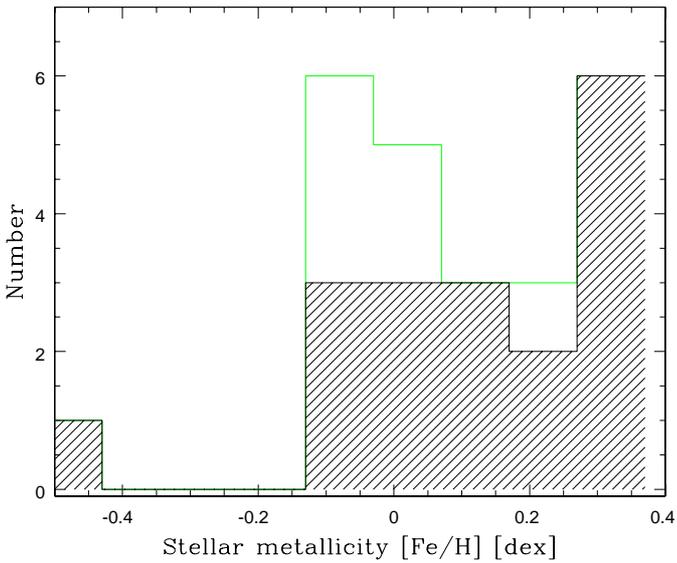}
      \caption{Histogram of host star metallicities [Fe/H] for the planets with semi-major axis greater than 4~AU (all are giant gaseous planets). In green, the histogram include the six candidates from Marmier et al. (in prep., private communication)
              }
         \label{metal}
   \end{figure}
%______________________________________________________________	

   Our targets are both bright (6.7$<$m$_{\mathrm V}$$<$7.6) and nearby (between 18 and 49~pc), hence ideal for follow-up surveys. The extension of the RV measurements for these targets will allow to refine the planetary parameters, to search for other planets in the systems, and to explore the magnetic activity of these stars.
 For orbital distances greater than 5~AU, imaging provide critical observational constraints on the system such as its inclination and enable to search for outer bodies or provide spectral information about the planet.
      The candidate planets would display astrometric signatures of hundreds of $\mu$as, for example, 550~$\mu$as on HD150706 and 175~$\mu$as on HD222155. Despite a duration mission of timescale shorter than the orbital period, part of these orbits should be easily detected by Gaia.
      Moreover, these systems with long-period low-eccentricity Jupiter-type planets may be similar to the solar system and contain lower mass planets in shorter orbits such as the $\mu$~Ara (HD160691) and 55~Cnc (HD75732) systems. New fiber scramblers were installed on SOPHIE in June 2011 (Perruchot et al. 2011), and preliminary tests showed that they provide a significant improvement in the stability of the spectrograph illumination, hence the RV accuracy. These stars will be followed-up with SOPHIE in order to search for multiplicity in these systems.
      
Hence, the transit probabilities for these candidates are very low at 0.07\% and 0.16\% for HD150706b and HD222155b, respectively. However, as they may host shorter-period low-mass planets with higher transit probabilities, they are good targets to search for Earth-like planets in transit around bright stars in order to identify a solar system twin.

\begin{acknowledgements}
      The authors thank all the staff of Haute-Provence Observatory for their 
contribution to the success of the ELODIE and SOPHIE projects and their support 
at the 1.93-m telescope. We thank the referee for his/her careful reading and judicious comments. 
We wish to thank the ``Programme National de Plan\'etologie'' (PNP) of CNRS/INSU, the 
Swiss National Science Foundation, and the French National Research Agency (ANR-08-JCJC-0102-01 and 
ANR-NT05-4-44463) for their continuous support of our planet-search 
programs. AE is supported by a fellowship for advanced researchers from the Swiss
National Science Foundation (grant PA00P2\_126150/1). 
IB and NCS would like to gratefully acknowledge the support of the European Research Council/European Community under the FP7 through a Starting Grant, as well from Funda\c{c}\~ao para a Ci\^encia e a Tecnologia (FCT), Portugal, through a Ci\^encia\,2007 contract funded by FCT/MCTES (Portugal) and POPH/FSE (EC), and in the form of grants reference PTDC/CTE-AST/098528/2008, PTDC/CTE-AST/098604/2008, and SFRH/BPD/81084/2011. DE and RFD are supported by CNES. This research has made use of the SIMBAD database and the VizieR catalog access tool operated at CDS, France. 
\end{acknowledgements}

%CAUP-11/2010-BPD, for the project FCOMP-01-0124-FEDER-009290&PTDC/CTE-AST/098528/2008

\appendix
%-------------------------
\section{Constraining the RV offset between ELODIE and SOPHIE}
\label{RVES}
When a star is observed by several instruments, the RV offsets between the different datasets are fitted as a free parameter in the Keplerian solution. A sample of about 200 stars, that had been selected as stable from ELODIE measurements, were also observed with SOPHIE to search for low-mass planets (Bouchy et al. 2009). This sample can be used to constrain the RV offset between the two spectrographs as these stars have a constant RV at the level of precision of ELODIE ($\sim$10ms$^{-1}$) on a timescale of several years. The $\Delta(\mathrm RV)$ is expected to depend on the color of the star ($B-V$) and to second order (that we neglect) on its metallicity.  Owing to its mean value of 0.003 given by Hipparcos, we neglect the error in the $B-V$. 

For both instruments, we compute the mean RV for each star, RV$_{ELODIE}$ and RV$_{SOPHIE}$. The error bars correspond to the quadratic sum of the standard deviations in the ELODIE and SOPHIE RV data. We then plot the difference $\Delta$(RV)$_{E-S}$=RV$_{ELODIE}$$-$RV$_{SOPHIE}$ as a function of the $B-V$ in Fig.~\ref{deltaES}. The RV$_{ELODIE}$ are shifted into the blue compared to RV$_{SOPHIE}$. We consider separately the RV measurements derived from the G2 (black squares) and the K5 (blue circles) cross-correlation mask. 

With the K5 mask, a linear fit (black dashed line) cannot be well-constrained and we choose a constant as an offset (green dashed line)
\begin{equation}
\Delta\mathrm{(RV)}_\mathrm{E-S} (K5)\, =\, -166\, \mathrm{ms}^{-1},
\end{equation}  
where the residuals have a dispersion of 20\,ms$^{-1}$, which is considered to be our error in the RV offset.

In the case of the G2 mask, stars with a $B-V$ $>$ 0.75 may have different properties from the others and should have been correlated with the K5 mask. This may be due to a bad spectral classification of these stars. We then fit a linear relation  considering only the stars with $B-V$ $<$ 0.75 (green line) 
\begin{equation}
\Delta\mathrm{(RV)}_\mathrm{E-S} (G2)\, =\, -425.6\,(B-V)\, + \,202.4\, \mathrm{ms}^{-1}.
\label{G22}
\end{equation}  
The residuals dispersion around the fit is 23 ms$^{-1}$, which we assume to be our offset calibration error. 

%------------------------------------------------------------------------------
\begin{figure}	
\center
\includegraphics[width=9cm]{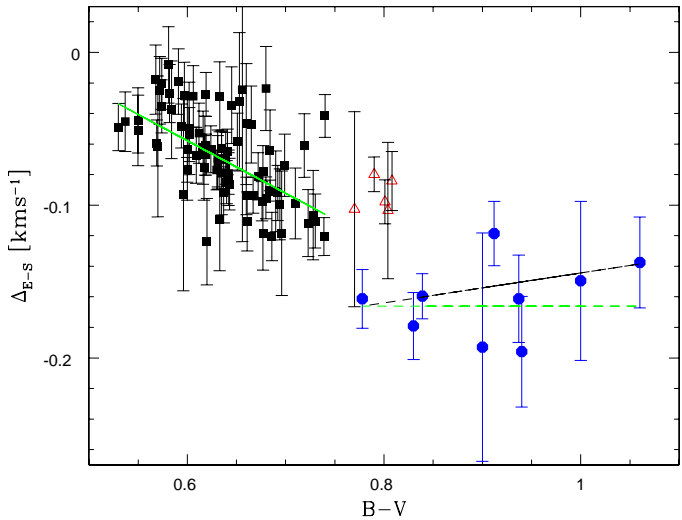}
\caption{Difference between the mean RV from ELODIE and from SOPHIE, $\Delta\mathrm{(RV)}_\mathrm{E-S}$ as a function of $B-V$ for a sample of stable stars. The error bars correspond to the quadratic sum of the standard deviation in the ELODIE and SOPHIE RV data. The green solid line is the best linear fit for stars correlated with a G2 mask (black squares). Those with $B-V$$>$0.75 are discarded (red triangles). The black dashed line is the best linear fit for stars correlated with a K5 mask (blue circles). The detection of the slope is insignificant and a constant value is chosen (green dashed line). }
\label{deltaES}
\end{figure}
%-------------------------------------------------------------------------------
% plot fait sous SP1/DELTARV.....
%-----------------------------------------

%------------------------------------------------------------------------------
\begin{figure}	
\center
\includegraphics[width=8cm]{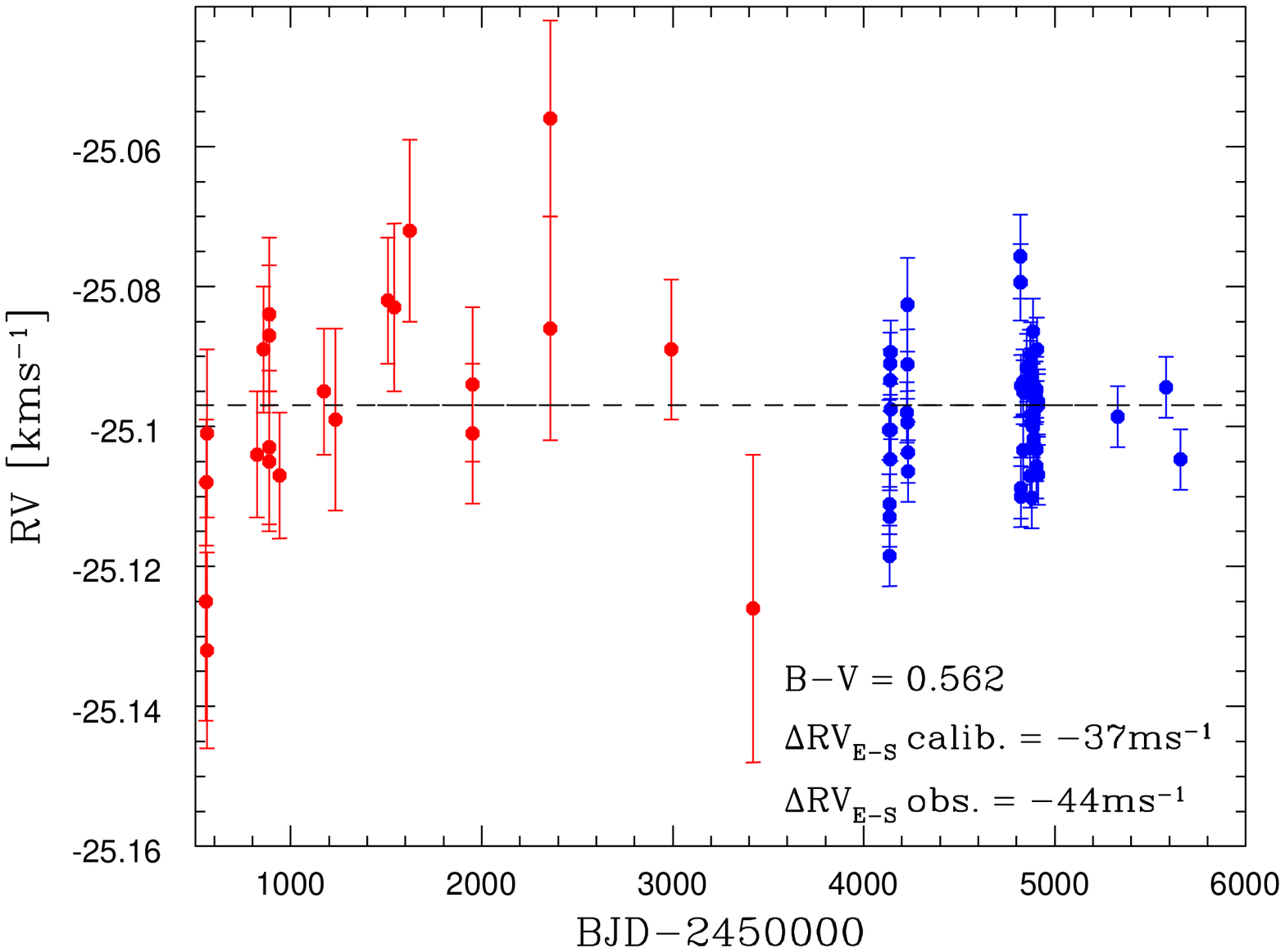}
\includegraphics[width=8cm]{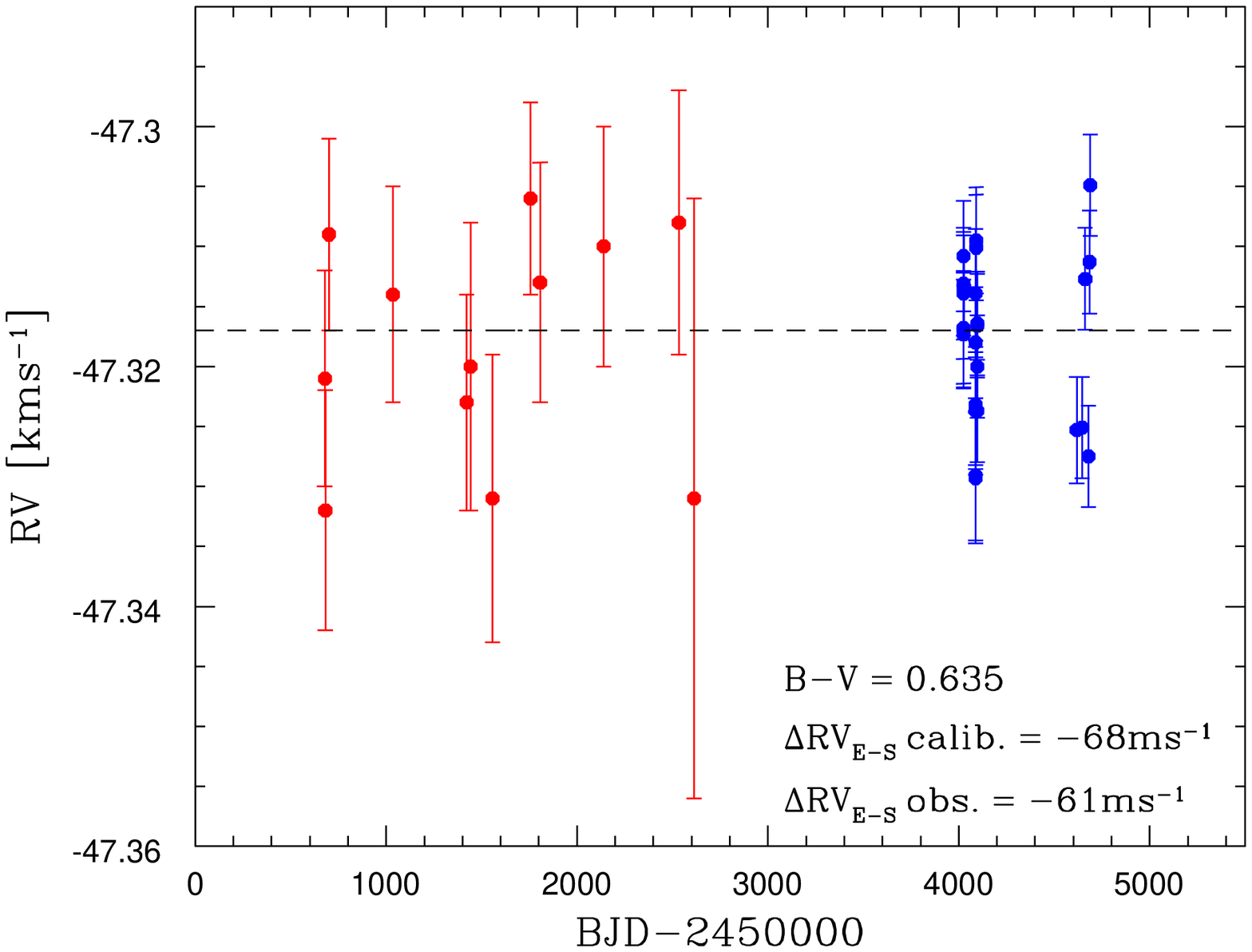}
\includegraphics[width=8cm]{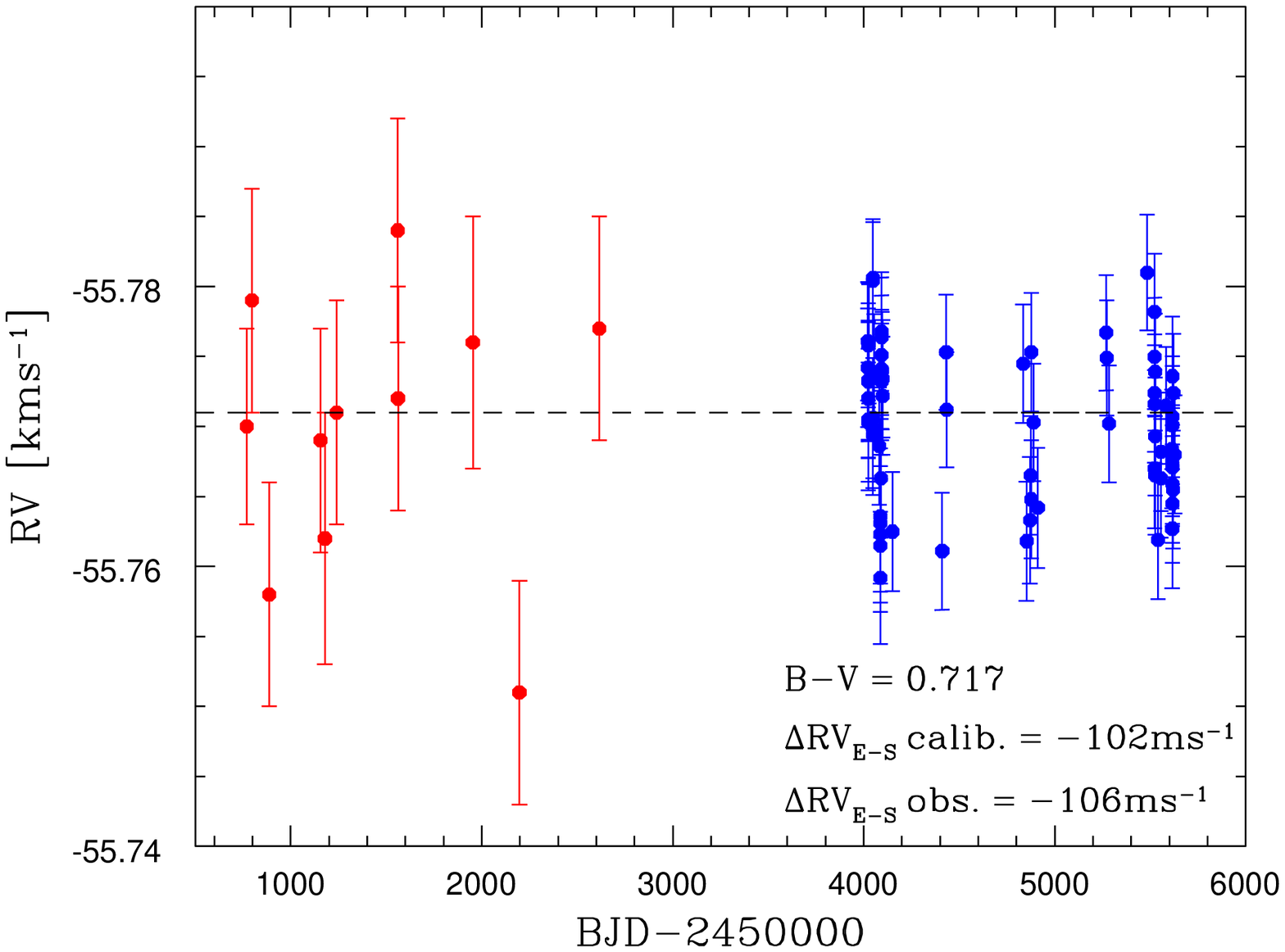}
\caption{ ELODIE (red points) and SOPHIE (blue points) RV of stable stars observed during more than 13 yr. 
 }
\label{deltaConstante}
\end{figure}
%-------------------------------------------------------------------------------

In Fig.~\ref{deltaConstante}, three stable stars observed over a period of more than 13~yr are shown, illustrating the reliability of the calibration.

\end{document}